\documentclass[10pt,preprint]{aastex}
\newcommand{\ee}[1]{\mbox{${} \times 10^{#1}$}}
\newcommand{\msun}{\mbox{M$_\odot$}}

\begin{document}

\title{Fragmentation and Evolution of Molecular Clouds. 
I: Algorithm and First Results}

\author{Hugo Martel\altaffilmark{1,2}, Neal J. Evans II,\altaffilmark{2}
and Paul R. Shapiro\altaffilmark{2}}

\altaffiltext{1}{D\'epartement de physique, g\'enie physique et optique,
Universit\'e Laval, Qu\'ebec, QC, G1K 7P4, Canada}

\altaffiltext{2}{Department of Astronomy, University of Texas, Austin, 
                 TX 78712}

\begin{abstract}
We present a series of simulations of the fragmentation of a
molecular cloud, leading to the formation of a cluster of protostellar cores.
We use Gaussian initial conditions with a power spectrum $P(k)\propto k^{-2}$,
assume an isothermal equation of state, and neglect turbulence and magnetic 
fields. The purpose of these simulations is to address a specific numerical 
problem called {\it artificial fragmentation}, that plagues simulations of
cloud fragmentation.
We argue that this is a serious problem that needs to be addressed,
and that the only reasonable and practical way to address it within
the Smoothed Particle Hydrodynamics algorithm (SPH) 
is to use a relatively new technique called {\it particle splitting}.

We performed three simulations, in which we allow $N_{\rm gen}=0$, 1, and 
2 levels of particle splitting. All simulations start up with $64^3$ SPH
particles, but their effective resolutions correspond to
$64^3$, $128^3$, and $256^3$ particles, respectively. The third simulation
properly resolves the Jeans mass throughout the entire system, at all times,
thus preventing artificial fragmentation.

The high resolution of our simulations results in the formation of a large
number of protostellar cores, nearly 3000 for the largest simulation.
This greatly exceeds the typical number of cores ($\sim60$) formed in
previous simulations, and enabled us to discover various processes that
affect the growth of the cores and the evolution of the cluster. 

The evolution of the cloud follows four distinct phases, or regimes.
Initially, during the {\it growth regime},
the cloud evolves into a network of intersecting filaments. After
roughly one dynamical time, core formation starts inside the dense 
gaseous fragments located at the intersection of the filaments, and to a
lesser extent inside the filaments themselves, and the
cloud enters the {\it collapse regime}. 
During this regime about 50\% of the gas, essentially the gas that started
up in overdense regions, is converted into cores. Competitive accretion
is the main process that controls the mass evolution of cores, but we
discovered that this process occurs {\it locally}, within each dense 
fragment. During the
following {\it accretion regime}, most of the remaining gas does not form
new cores, but rather accretes onto the existing cores. Eventually each
gaseous fragment has turned into a subcluster of cores, and
these subclusters later merge to form the final cluster. The gas
left in the system has become negligible, and the system has
reached the {\it N-body regime}, in which the dynamics of the cluster is
governed by N-body dynamics.

The final mass distribution of cores has a lognormal distribution,
whose mean value is resolution-dependent; the distribution shifts down
in mass as the resolution improves.  The width of the
distribution is about 1.5 (e.g., a factor of 30 in the mass), and
the low-mass edge of the distribution corresponds to the lowest
core mass that the code can resolve. This result differs from previous claims
of a relationship between the mean of
the distribution and the initial Jeans mass.
 
\end{abstract}


\section{Introduction}

\subsection{Clustered Star Formation}

An understanding of star formation is pivotal to understanding both
the origin of galaxies and the origin of planetary systems. 
A crucial step in the formation of galaxies is the formation of stars. 
Intensive archaeology of chemical abundances has uncovered the outline
of the star formation history of our own galaxy (e.g., \citealt{mcwilliam97}), 
and look-back studies are beginning to provide more direct information about
star formation beyond $z \sim 1$. 
It now appears that the star formation rate was flat to $z >3$
\citep{madau99,smailetal00} or even increasing up the highest observed
redshifts \citep{lanzettaetal02}.
At least half the extragalactic background light lies in the far-infrared
to submillimeter region \citep{pugetetal96,hauseretal98}, suggesting
that half the star formation in the history of the Universe has
occurred in dusty regions. Submillimeter galaxies, at a mean
$z = 2.3$ \citep{chapmanetal05}, are forming stars at 
prodigious rates (up to 1000 solar masses/year), in dusty, molecule-rich 
environments (e.g., \citealt{bargeretal98,cimattietal98,coxetal02}).
Such objects have now been seen even at $z = 6.4$ \citep{walteretal04}.
To understand the nature of this process, we must develop a better 
understanding of massive star formation in similar environments more
amenable to detailed study, the dense regions of molecular
clouds in our Galaxy. 

In a similar way, the study of the origin of planetary systems rests
on a deeper understanding of the role of disks in star formation. It is
now clear that disks surround most young stars \citep{beckwithetal90}, and
we are beginning to study the properties and evolution of these disks
(e.g., \citealt{mlw00}). 
These studies are fundamental to understanding the frequency
and variety of planetary systems that current searches are revealing
(e.g., \citealt{cmb99,jmu01,marcy05}).

Over the last two decades, we have made substantial progress in 
understanding how low-mass stars form in relative isolation.
Spurred by the elaboration of an elegant theoretical paradigm,
beginning with \citet{shu77} and culminating in the influential review
by \citet{sal87}, observers have developed the 
capability to test predictions of theoretical models. We have 
a ``standard model'' and a number of variations, each of which makes
predictions that can be tested by observations.
In particular, the
form of the density and velocity field in the envelope around the
forming protostar differs in the different models, and observers are
beginning to be able to distinguish these observationally (see reviews
by \citealt{evans99,meo00,awtb00}).
Studies of dust continuum emission with instruments like {\sl SCUBA} is 
providing a valuable new probe of the density distribution, while 
molecular line profiles probe the kinematics (e.g.,
\citealt{zhouetal93,ze94,meo00}). 
In addition, predictions that disks would form on scales of $1-100$~AU
around forming stars (e.g., \citealt{tsc84}) have been
verified \citep{beckwithetal90}.
In the study of low-mass, isolated star formation, theory has clearly
revealed the path to the observers.

However, most stars probably form not in isolation, but in clustered
environments \citep{elmegreen85,cbh99,evans99,pf99,elmegreenetal00,ll03}.
Massive stars seem to form exclusively in these environments, but
the full spectrum of stars and sub-stellar objects forms in 
clusters \citep{ll95}. In addition,
the bursts of star formation seen in distant galaxies are clearly
related to the formation of massive stars in clusters.
Consequently, understanding them is crucial to 
understanding galaxy formation. Disks around forming stars have been
seen in clustered environments (e.g., \citealt{odw94,strom95}), 
but interactions among proximate star/disk systems may
affect the mass distribution of these disks \citep{eisner03}.

In the area of massive, clustered star formation, 
extensive observations exist (e.g., \citealt{churchwell93,kurtzetal00}), 
but theory is very underdeveloped. We know that the
typical conditions in dense regions of high mass are quite different from those
in dense regions of low mass (e.g., \citealt{plumeetal97,shirleyetal03b}). 
The temperatures and densities are higher, and the
masses of gas are sufficient to form many stars (up to $10^4M_\odot$).
The linewidths of molecular lines are much 
greater and deviate from the linewidth-size relations known from lower-mass
regions and regions not forming stars \citep{shirleyetal03a}. 
The overall density distribution in these high-mass regions 
seems well approximated 
by a power law $n(r) = n_f (r/r_f)^{-p}$,
with exponent similar to that of low-mass regions ($p\sim1.8$)
\citep{shirleyetal03b}
but densities ($n_f$) 100 times higher \citep{evansetal02,muelleretal02}.
These wide lines are highly
supersonic, and turbulence must play a major role. Some dense regions seem
to be more fragmented \citep{wangetal93,lef97}, 
as might be expected if clusters are forming.

In these conditions, direct extension of theories of low mass star
formation may run into problems. Recent work on the formation of massive
stars delineates the theoretical framework \citep{mkt02,mkt03} of formation
in turbulent dense regions. 
This work predicts the overall properties of the regions,
but the details of the formation and its observational manifestations
remain to be understood. The close proximity of
other clumps within the overall dense region 
will perturb the density and velocity fields around a given
clump, making the kinds of tests applied to isolated regions less
meaningful (and very difficult observationally). Theories that make
detailed predictions of observables are needed. Statistical measures must
be compared between theory and observation, rather than any specific
realization of a simulation because stochasticity is inevitable.
For example, the
distribution of clump masses and velocities as a function of time
could be predicted and compared to observations. If one can follow the
process with sufficient dynamic range, the distribution of clump angular
momenta could be used to predict things like the frequency of binaries
and disks, if supplemented by more detailed calculations of the subsequent
evolution of individual structures once they get small enough that their
internal dynamical time becomes less than the interaction time. Indeed,
predictions of when that point occurs are needed. Larger scale correlations
can be studied, such as the tendency toward alignment of angular momentum
vectors or magnetic fields.

A number of simulations of the fragmentation of a molecular cloud to form 
a cluster have been performed 
(e.g., \citealt{larson78,klm91,bbcp97,bcl99,pn02,tp04}; see
the recent review articles by \citealt{larson03,mk04}). 
Two research groups have been particularly prominent in recent years.
Klessen, Burkert, and
their collaborators (\citealt{kbb98,kb00,kb01,khml00,klessen01a,klessen01b,
sk04,jk04}, hereafter collectively KB;
\citealt{lkm03,jappsenetal05}),
have used a Smoothed Particle Hydrodynamics (SPH) code to 
follow the evolution of a region with many times the Jeans mass. 
They were able
to reproduce the observed distribution of clump masses, but found that 
bound units developed a steeper distribution, more similar to that of
stars. This result meshed nicely with recent observations showing that
the densest structures in several clouds also have a steeper mass function
for masses above $0.5\rm M_\odot$ \citep{man98,ts98}.
Bonnell, Bate, and their collaborators constitute the second
group \citep{bbb02a,bbb02b,bbb03,bb05,daleetal05,dbc05}. They also used
SPH, but have focused mostly
on smaller, denser systems for which the
gas is optically thick at high densities and the assumption
of isothermality breaks down.

An important issue in such simulations is the phenomenon called
artificial fragmentation, which can lead to severe numerical problems.
In the next subsection, we describe the approach used by the
various groups for dealing with this problem.

\subsection{Artificial Fragmentation and the Jeans Criterion}

\citet{trueloveetal97} and \citet{boss98}
have shown that a minimum requirement of any
grid-based simulation of a fragmenting cloud is that the algorithm
can resolve the Jeans mass $M_J$ (we shall refer to this as the 
{\it Jeans criterion}).
The maximum mass inside a cell must be smaller than $\sim1/64$ of
the Jeans mass in order to prevent a spurious, resolution-dependent 
effect they called {\it artificial fragmentation}. This effect, 
if present, can invalidate the results of star-formation simulations, 
by producing initial mass functions and accretion histories that are
totally wrong.
\citet{bb97}
derived a different, but
equivalent criterion for SPH. The total mass contained inside
the zone of influence of a particle must be less than about twice the
Jeans mass in order to prevent artificial fragmentation.
As \citet{kleinetal99} pointed out, this requirement poses a serious problem
for isothermal simulations performed with SPH. The Jeans mass varies as
$M_J\propto T^{3/2}\rho^{-1/2}$. Hence, in
isothermal clouds, $M_J\propto\rho^{-1/2}$, so 
the Jeans mass decreases as collapse proceeds. 
Since the smoothing lengths $h$ are
adjusted in such a way that the mass inside the
zone of influence of every
particle remains constant, the Jeans mass will eventually be
underresolved as the density increases.

There is, however, a physical lower limit to the mass of fragments, 
simply because the isothermal approximation breaks down at
sufficiently high densities. \citet{bbb97}, \citet{bate98}, and
\citet{kleinetal99}
extend their simulations into the high density regime by using a
barotropic equation of state, which is isothermal below a certain
critical density $\rho_{c2}$, and adiabatic above it
(the actual equation of state is significantly more complex 
[see, e.g., \citealt{scaloetal98}], but the barotropic form is a convenient
approximation).
In the adiabatic regime, $P\propto\rho^{5/3}$,
hence $T\propto\rho^{2/3}$ and therefore $M_J\propto\rho^{1/2}$. 
There is therefore a minimum Jeans mass, corresponding to
the critical density $\rho_{c2}$.
As long as the Jeans criterion is satisfied at that density, it will
be satisfied in the entire system, at all times.

This still poses a problem. The isothermal approximation is valid
for densities up to $10^{10}\,\rm cm^{-3}$ \citep{kb00}.
For a region with mean density $10^2\,\rm cm^{-3}$, 
the range of $10^8$ in density corresponds to
a range of $10^4$ in Jeans mass. Since the smallest Jeans mass must contain 
$\sim100$ particles to satisfy the Jeans criterion, {\it the total
number of particles in the simulation must be at least 1
million}, and this would satisfy the Jeans criterion only marginally.

Klessen, Burkert, and their collaborators do not address the problem
of artificial fragmentation. They are aware of the problem, but believe
that this problem does not affect their conclusions significantly.
Bate, Bonnell, and their collaborators are very concerned with
artificial fragmentation. They avoid the problem by simulating low-mass
systems ($M\sim50\rm M_\odot$), so that their resolution is
sufficiently large to reach the regime where the gas becomes adiabatic.
They can then use a barotropic equation of state.

In this paper, we consider an alternative approach that can
simulate high-mass systems, in the isothermal regime, while still
solving the artificial fragmentation problem.
This is achieved by using {\it particle splitting},
a relatively new and very promising technique \citep{kw02}. The basic idea
consists of starting the simulations with a manageable number of
SPH particles, and then refining the mass resolution {\it locally\/} in regions
where additional resolution is needed
to satisfy the Jeans criterion. Original particles are automatically
replaced (or ``split'') by a more finely-spaced set of smaller-mass
particles wherever extra resolution is required.
This can be seen as the Lagrangian
counterpart of the Adaptive Mesh Refinement techniques used in Eulerian,
grid-based algorithm. In our implementation of particle splitting,
SPH particles split into 8 equal-mass particles when the Jeans
criterion is locally violated. For an isothermal gas, this results
in a new generation of particle splitting every time the density
increases by an additional factor of 64. 

\subsection{Objectives}

The primary goal of this ongoing
project is to study the effect of feedback from clustered star formation on
the evolution of the ISM. However, the issue of feedback will not be
addressed in this first paper, for the following reason: The existence of
artificial fragmentation casts a huge shadow over all
SPH simulations of cloud fragmentation that assume an isothermal equation of
state, raising doubts about the validity of such simulations. We feel
that this problem is far too important to be ignored, and must
be addressed first, and successfully, before we even consider implementing
additional physical processes into the algorithm. {\it Solving the problem
of artificial fragmentation is an essential first step}.
In this paper, we address this problem using a SPH algorithm which combines
self-gravity, hydrodynamics, particle splitting, and sink particles. This
algorithm is described in \S2 below.

The main objectives of this paper are the following:

1. {\it Test the feasibility of particle splitting, and investigate, both
analytically and numerically, the interplay between particle
splitting, sink particles, and the Jeans criterion.} Using particle splitting
enables us to start up simulations with a small number of particles for a
given resolution. However, the feasibility of this approach depends critically
on the efficiency of particle removal by sink formation and accretion onto
sinks. If a large number of particles split before the formation and growth
of sinks becomes important, the total number of particles in the simulation
might become too large to be manageable. It is not obvious a priori that
sink formation and growth will remove particles sufficiently rapidly to
offset the increase in particle number resulting from splitting. One of
the main goals of this paper is to determine if the peak number of particles
in such simulations remains manageable.

2. {\it Perform a convergence study.}
We can ascribe
to each simulation with particle splitting an ``effective particle number,''
which is the number of particles a simulation without particle splitting
would need to achieve the same resolution in dense regions. The three
simulations presented in this paper start with $64^3$ particles, and allow
for $N_{\rm gen}=0$, 1, and 2 generations of particle splitting, respectively.
The splitting factor is $f_{\rm split}=8$, meaning that when a particle
splits, it is replaced by 8 particles, each having 1/8 of the mass of
the parent particle. Hence, the effective particle numbers of the three 
simulations are $64^3$, $128^3$, and
$256^3$, respectively. Since we are using identical
initial conditions, these three simulations taken together constitute
a convergence study, the largest one ever performed for such
simulations.

3. {\it Perform the largest simulation of this kind ever done,
in terms of effective number of SPH particles or number of protostellar
cores formed.}
For our largest simulation ($N_{\rm gen}=2$), the effective particle
number is $256^3=16,777,216$, about 33 times the largest number of
particles used by KB ($500,000$). Our smallest simulation ($N_{\rm gen}=0$,
that is, no particle splitting) uses $64^3=262,144$ particles, which
is comparable to the largest isothermal
simulations of KB. Furthermore, we start
the simulations with $N_J=500$ Jeans masses instead of KB's 222 Jeans
masses. As a result, we will form a much larger number of protostellar
cores. This has three advantages: First, our determination of the
initial mass function of protostellar cores will be more accurate.
Second, it will enable us to study the mass assembly
history and final structure of the cluster in more detail. And third,
forming a larger number of cores might lead to the discovery of some
interesting processes in the evolution of the cluster, that would not
occur in a cluster with much fewer cores; as we shall see, this is
indeed the case.

\section{THE NUMERICAL ALGORITHM}

\subsection{Basic Equations}

The evolution of a self-gravitating gas is described by the
conservation equations for mass, momentum, and energy, coupled with
the Poisson equation and the equation of state,
\begin{eqnarray}
&&{\partial\rho\over\partial t}+\nabla\cdot(\rho{\bf v})=0\,,\\
&&{\partial{\bf v}\over\partial t}+({\bf v}\cdot\nabla){\bf v}=
-{\nabla P\over\rho}-\nabla\phi\,,\\
&&{\partial\epsilon\over\partial t}+{\bf v}\cdot\nabla\epsilon=
-P\nabla\cdot{\bf v}+{\Gamma\over\rho}-{\Lambda\over\rho}\,,\\
&&\nabla^2\phi=4\pi G(\rho-\bar\rho)\,,\\
&&P=f(\rho,\epsilon)\,,
\end{eqnarray}

\noindent
where $\rho$ is the density, $P$ is the pressure,
$\epsilon$ is the
specific internal energy, ${\bf v}$ is the velocity,
$\phi$ is the gravitational potential, $\bar\rho$ is the mean density,
and $\Gamma$ and $\Lambda$ are the radiative heating and cooling rates,
respectively. Equation~(4) requires some explanation. To prevent the
overall collapse of the cloud, we assume that the cloud is essentially 
infinite, and use periodic boundary conditions. 
However, a periodic gravitational
potential $\phi$ is only possible if the total mass of the system vanishes.
By adding the term $-\bar\rho$ in equation~(4), the effective mass, defined
as the integral of the source term $\rho-\bar\rho$ over the computational
volume, does vanish, and a periodic solution for $\phi$ becomes possible.
A valid interpretation of equation~(4) is that the term $-\bar\rho$
accounts for whichever process makes the cloud
{\it globally\/} stable, while it is the fluctuation $\rho-\bar\rho$
that make the cloud {\it locally\/} unstable.
\citet{khml00} and \citet{mk04} have suggested that supersonic turbulence
might explain the global stability of clouds.

It is important to realize that equation~(4) is still physically correct.
Since the cloud is assumed to be 
infinite, the term $-\bar\rho$ represents a uniform, negative density 
background extending to infinity in all directions, and such component cannot,
by symmetry, exert any force in any direction on a mass element.
For a more formal description, we refer the reader
to \citet{al98}, and references therein.

In this paper, the set of equations we are using is significantly simpler.
First, since we use Smoothed Particle Hydrodynamics (SPH), 
a Lagrangian, particle-based algorithm, mass is automatically conserved,
and we can ignore the continuity equation~(1). Second, we assume that the
gas is isothermal. The specific internal energy $\epsilon$ is therefore
constant in space and time, and equation~(3) can be ignored as well. 
The equation of state, equation~(5), becomes
\begin{equation}
P={c_s^2\rho\over\gamma}\,,
\end{equation}

\noindent where $c_s(\epsilon)$ is the sound speed, which is constant
in an isothermal gas, and $\gamma$ is the polytropic constant\footnote{
The relationships between the concepts of polytropic constant, polytropic
equation of state, and isothermality are often reported incorrectly.
The polytropic constant $\gamma$ is the ratio of the specific heat
capacities at constant pressure and constant volume. A polytropic equation
of state has the form $P\propto\rho^\gamma$, but such an equation is valid
only if the entropy of the gas is constant both in space and time. It is often
said that $\gamma=1$ for an isothermal gas, but this is incorrect in general.
In our simulations, $\gamma=5/3$, the equation of state is {\it not\/}
polytropic, and it is the coupling between the
gas and a background radiation field that makes the gas isothermal.
}.
The system of equations (1)--(5) reduces to
\begin{eqnarray}
&&{d{\bf v}\over dt}=-{c_s^2\over\gamma\rho}\nabla\rho-\nabla\phi\,,\\
&&\nabla^2\phi=4\pi G(\rho-\bar\rho)\,,
\end{eqnarray}

\noindent where 
$d/dt\equiv\partial/\partial t+{\bf v}\cdot\nabla$ is the Lagrangian
time derivative.

\subsection{The $\bf SPH/P^3M$ Algorithm}

We use a hybrid gravity/hydrodynamics algorithm.
The gravitational forces are computed using a $\rm P^3M$ algorithm
\citep{he81}, while the gasdynamical equations are solved using 
the SPH algorithm (see Monaghan 1992, and references therein). 
This hybrid algorithm was originally introduced by \citet{evrard88},
though we have developed our own version. Our code
is actually an Adaptive SPH (ASPH) code \citep{shapiroetal96,owenetal98}.
The ASPH algorithm uses anisotropic smoothing kernels, and a special
treatment of artificial viscosity that prevents spurious preheating
of low-density regions. However, in this paper, we consider an isothermal
equation of state, so preheating is not an issue. We decided to use
isotropic smoothing kernels for now (that is, using the Adaptive SPH code 
as a standard SPH code). We will consider anisotropic smoothing kernels in
future work.

We have modified our original algorithm to include a treatment
of particle splitting and sink particles. The implementation
of these features is described below.

\subsection{Particle Splitting}

We have modified our original SPH algorithm to include
particle splitting and sink particles. Particle splitting
is implemented as follows.
We assume that there is a minimum number of particles
$n_{J,{\rm a.f.}}$ that a Jeans mass must contain in order
to be properly resolved and not undergo artificial fragmentation.
This gives us the {\it Jeans criterion}. Each particle $i$
is required to satisfy the condition
\begin{equation}
{M_J(\rho_i,\epsilon_i)\over m_i}\geq n_{J,{\rm a.f.}}\,,
\end{equation}

\noindent where $M_J$ is the Jeans mass, which is a function of
the density $\rho$ and specific internal energy $\epsilon$,
$M_J(\rho_i,\epsilon_i)$ is the value of $M_J$ evaluated
at the location of particle $i$ (see \S4.3 below), and $m_i$ is
the mass of particle $i$.
When this condition is violated, the algorithm responds by
splitting particle $i$ into $f_{\rm split}$ particles of
mass $m=m_i/f_{\rm split}$. These new particles might themselves be
split later as $M_J$ keeps decreasing. For instance, in the case of
an isothermal calculation ($\epsilon=\hbox{const}$), 
$M_J\propto\rho^{-1/2}$, and there is a series of particular densities
$\rho=a\bar\rho$, $a\bar\rho f_{\rm split}^2$, $a\bar\rho f_{\rm split}^4$,
$\ldots$, ($a=\hbox{const}$) at which particle splitting will occur.

When a particle is split, the algorithm must determine the positions,
velocities, masses, and smoothing lengths of the daughter 
particles\footnote{and also the specific internal energies, if the gas is not 
isothermal.}.
In the original approach of \citet{kw02}, 
particles are split into a sphere
of 13 daughter particles forming a compact lattice. We use a different 
approach,
in which particles are split into $f_{\rm split}=8$ particles located
on the vertices of a cube. This approach will be
easier to generalize to Adaptive SPH later.
We create the daughter particles as follows:
Consider a particle $i$, with position ${\bf r}_i$, velocity ${\bf v}_i$,
mass $m_i$, and smoothing length $h_i$, that violates the condition~(9), and
therefore needs to be split.
If $(\Delta r)_i$ is the mean particle
spacing in the vicinity of particle $i$, the 8 daughter particles will
be located at
\begin{equation}
{\bf r}={\bf r}_i+{(\Delta r)_i\over4}\left[\matrix{
\pm1\cr\pm1\cr\pm1\cr}\right]\,,
\end{equation}

\noindent so that the spacing between daughter particles will be
$(\Delta r)_i/2$.\footnote{Note: because of the periodic boundary conditions,
some daughter particles are wrapped around the computational box if particle
$i$ is located less than $(\Delta r)_i/4$ away from the edge of the
box.} In the initial conditions, the smoothing lengths $h_i$ of the particles
are initialized to be a multiple of the mean particle spacing:
$h_{i,{\rm init}}=\ell_2\Delta r$, where $\ell_2$ is a constant (using
the notation of \citealt{shapiroetal96}). Then, as the calculation
proceeds, the algorithm evolves the smoothing
lengths $h_i$ is such a way that this relation is (roughly) maintained
locally. Hence, the local particle spacing can be estimated using
$(\Delta r)_i=h_i/\ell_2$. Equation~(10) reduces to
\begin{equation}
{\bf r}={\bf r}_i+{h_i\over4\ell_2}\left[\matrix{
\pm1\cr\pm1\cr\pm1\cr}\right]\,.
\end{equation}

\noindent In our simulations, we generated initial conditions using 
$\ell_2=2$, to provide a sufficient number of neighbors.

We set the velocity of the daughter particles equal to ${\bf v}_i$.
We could use a more sophisticated approach that would take 
the local velocity gradient into account, but this is not really necessary. 
Particles tend to readjust themselves in one time step, erasing any
velocity fluctuation at scales smaller that the smoothing length.
Finally, we
set the masses of the daughter particles equal to $m_i/8$, and their
smoothing lengths equal to $h_i/2$.

\subsection{Sink Particles}

Sink particles must be used in simulations of cloud fragmentation to reduce
the timesteps, if
we hope to determine the initial mass fraction of collapsed fragments.
It is not sufficient to use an algorithm with individual timesteps,
because as the gas fragments and collapses, most SPH gas particles end up
in dense regions with short dynamical times. By replacing each dense clump
of gas particles by a single object, called a sink particle
\citep{bbp95,bcl02}, we can
eliminate this problem and increase the speed of the algorithm 
tremendously. In our implementation of sink particles, we use the method of 
\citet{bbp95}, and we refer the reader to that paper for details.

\subsubsection{Creation of Sinks}

Our algorithm uses individual timesteps. Each particle $i$ is given a timestep
$\Delta t_i=(\Delta t)_{\rm basic}/2^n$, where $(\Delta t)_{\rm basic}$
is the ``basic timestep,'' and $n$ identifies the timestep bin where the
particle belongs.
Sink particles are created only at the end of the basic timesteps,
when all the particles are in sync. 
This greatly simplifies the implementation.
Gas particles are replaced by sinks when they reach  
a threshold density $\rho_c$. This density
threshold is a numerical parameter that does not have any physical
meaning, except for the fact that the assumption of isothermality
must be physically valid at all densities below $\rho_c$. The smallest
collapsed objects that can form in the simulations will have a mass
equal to $(M_J)_c$ the Jeans mass at the density $\rho_c$. Hence, $\rho_c$ 
fixes the mass resolution of the algorithm, much in the same way as the 
softening length fixes its length resolution.

To create sinks, the algorithm identifies all particles
whose density\footnote{We use the expression ``density of a particle''
for convenience; strictly speaking, SPH particles do not carry a
density, hence the correct expression is ``the density of the gas at
the location of a particle.''}
exceeds the threshold density $\rho_c$ for
sink creation. These particles are then sorted in decreasing order
of the density. Starting with the densest particle, the algorithm finds
all $n_{\rm acc}$ particles within a fixed radius $r_{\rm acc}$ of that
densest particle, where $r_{\rm acc}$ is
called the accretion radius.
An important issue is how to determine the
appropriate value for $r_{\rm acc}$, or equivalently, how many
particles $n_{\rm acc}$ should be turned into a sink.
We shall assume that it takes a
minimum number of particles, $n_{J,\min}$, to properly resolve
a Jeans mass, and we will set that number equal to the 
number of particles $n_{J,{\rm a.f.}}$ that a Jeans mass must contain to
prevent artificial fragmentation. This makes sense, since the negative 
consequence of underresolving the Jeans mass is precisely to cause
artificial fragmentation.
The number of particles $n_{\rm acc}$ must also exceed the number of
particles $(n_J)_c$ inside a Jeans mass at the threshold density $\rho_c$,
otherwise sub-Jeans mass objects would be turned into sinks. Hence, the
condition is
\begin{equation}
n_{\rm acc}\geq\max\left[n_{J,\min},(n_J)_c\right]\,.
\end{equation}

\noindent 
This ensures that every fragment replaced by a sink particle (i) is
properly resolved, and (ii) contains at least a Jeans mass.
We adjust $r_{\rm acc}$ such that 
$n_{\rm acc}\approx\max\left[n_{J,\min},(n_J)_c\right]$
at the threshold density $\rho_c$. If the values of $n_{\rm acc}$
are systematically smaller than $\max\left[n_{J,\min},(n_J)_c\right]$,
this would imply that we chose a value of $r_{\rm acc}$ that was too small,
and sinks would then form at densities much larger than $\rho_c$.
Conversely, if the values of $n_{\rm acc}$ are systematically
larger than $\max\left[n_{J,\min},(n_J)_c\right]$, this would indicate
a value of $r_{\rm acc}$ that was too large, effectively reducing
the mass resolution of the algorithm, by forming objects that
are too massive. The actual determination of $r_{\rm acc}$ is described below.

To create a sink particle, a second condition must be satisfied,
\begin{equation}
\label{alpha}
\alpha\equiv{E_{\rm th}\over|E_{\rm gr}|}<1\,,
\end{equation}

\noindent where $E_{\rm th}$ and $E_{\rm gr}$ are the thermal and
gravitational energies of the particles inside the accretion radius,
respectively. \citet{bbp95} 
use $\alpha<0.5$ as a criterion. However, using equation~(\ref{alpha})
is more consistent with the fact that we are trying to turn Jeans-mass
clumps into sinks, since by definition $\alpha=1$ for a uniform sphere
of mass $M=M_J$. If both conditions are satisfied, the particles inside
$r_{\rm acc}$ are removed and replaced by a sink particle, which
inherits the properties of the parent particles (center of mass position and
velocity, total mass, and total angular momentum). The
algorithm then selects the next densest particle {\it still available}
(some particles with $\rho>\rho_c$ might have been incorporated into
sinks created around denser particles), and the process is repeated
until all particles with $\rho>\rho_c$ have been considered.

\subsubsection{Mass Evolution of Sinks}

Two physical processes can increase the mass of protostellar cores: 
mergers with other cores, and accretion of gas by cores. In the algorithm, 
these processes correspond to merging of sinks, and accretion of SPH gas 
particles by sinks, respectively.

The merging of sinks is an interesting issue. Allowing sinks to merge would
clearly affect the final mass distribution of protostellar cores.
Unfortunately, no known prescription exists for implementing
sink merging into a SPH algorithm. We intend to investigate this issue
in a separate paper, but for now we will make the usual assumption that
sink particles do not merge, as in \citet{bbp95}, KB, and others.

Accretion of gas particles onto sinks is also performed at the end of
each basic timestep. The algorithm checks for gas particles that are located
inside the accretion radius $r_{\rm acc}$ of a sink particle. These
particles are then accreted by the sinks, and the sink properties
(position, velocity, mass, angular momentum) are updated accordingly.
If a gas particle is within the accretion radius of several sinks, we
compute the total energy of the (particle + sink) systems, and choose the sink
for which this energy is the smallest. Unlike 
\citet{bbp95}, we do not require that
the energy be negative. Such a criterion would fail in general, because in
these simulations the gas tends to fall onto sinks at supersonic speeds
that exceed the escape velocity from the sink. This happens when gas particles
are accelerated by large mass concentrations, that might contain several sinks.
A gas particle approaching a cluster of sinks will be accelerated by
the whole cluster, and will acquire a velocity comparable to the escape
velocity from the whole cluster. But this velocity will always exceed 
the escape
velocity from the particular sink that will accrete that particle.
Hence, the gas particle will not be gravitationally bound to the sink
onto which it accretes. This is not a concern, because the physical
process responsible for the final stage of accretion is not
gravitational capture, but rather physical collision between a gas particle
and a sink. In a simulation without sinks, 
a gas particle approaching a dense clump at supersonic
speed would be decelerated down to subsonic speeds by the artificial
viscosity, resulting in the conversion of kinetic energy into thermal
energy, that would then be radiated away since the gas is
assumed to be isothermal. But this process would occur at scales
smaller than $r_{\rm acc}$, and therefore cannot occur when clumps are
being replaced by sinks. By allowing large-velocity particles to accrete
onto sinks, we are essentially putting-in the subgrid physics of collision,
viscous
deceleration, and radiative dissipation by clumps smaller than $r_{\rm acc}$.

When using sinks, boundary conditions must be applied at the accretion
radius $r_{\rm acc}$, otherwise the SPH calculation of the density,
pressure forces, and viscous forces on gas particles located
immediately outside $r_{\rm acc}$ would be incorrect, resulting
in spurious effects. This is discussed in great detail in \citet{bbp95},
and also in \cite{bcl02}. Our implementation of boundary conditions
follows the description of \citet{bbp95}. Interestingly, in our simulations,
the effect of boundary conditions turns out to be quite small. We found
that most gas particles fall radially toward sinks, and collide with
them at supersonic speeds. As \citet{bbp95} point out, viscous
boundary conditions have a negligible effect for particles falling
radially, while pressure boundary conditions have a negligible effect
for particles moving supersonically. Our findings are consistent with that
claim.

\section{CLOUD FRAGMENTATION}

In this section, we discuss some aspects of the numerical
simulations of cloud fragmentation, including the interplay between resolution,
artificial fragmentation, particle splitting, and sink pyarticles.
For the sake of the discussion, we consider a particular simulation of KB,
which we call the ``benchmark simulation.'' For this particular simulation,
the number of particles is $N=200,000$, the initial number of Jeans masses
is $N_J=222$, the mean density, in computational units,
is $\bar\rho=1/8$, and the threshold density for creating sinks is
$\rho_c=40,000\bar\rho$. For the number of particles necessary to prevent
artificial fragmentation, we assume $n_{J,\rm a.f.}\approx100$ (this 
is a guess; no
value of $n_{J,\rm a.f.}$ is provided by KB). Notice that this is one of
the highest-resolution simulation performed by KB. They have performed
a higher-resolution simulation with $N=500,000$, and by lowering the
threshold density and the initial number of Jeans masses, they were able
to increase their mass resolution significantly. Because of the reduced
number of Jeans masses in this simulation, the more suitable calculation
for us to use as a benchmark is the one with $N=200,000$.

\subsection{Mass Resolution}

We can estimate the effect of particle splitting on the mass resolution
of the algorithm using a simple calculation. We assume that the system
has a total mass $M_{\rm tot}$, and the simulation starts up with $N$
equal-mass particles of mass $m=M_{\rm tot}/N$. Because the initial
density fluctuations are small ($\rho_{\rm init}\approx\bar\rho$),
and the gas is isothermal,
the initial Jeans mass $M_{J,{\rm init}}$ is
essentially constant in space. The initial number of Jeans masses in
the system is then given by $N_J=M_{\rm tot}/M_{J,{\rm init}}$, and
the initial number of particles in each Jeans mass is given by
$n_{J,\rm init}=N/N_J=M_{J,{\rm init}}/m$.

In the initial state, the system must satisfy the following conditions:
\begin{eqnarray}
\label{condaf}
n_J&>&n_{J,\rm a.f.}\,,\\
\label{condc}
\rho&<&\rho_c\,.
\end{eqnarray}

\noindent
The first condition states that the Jeans mass is properly resolved initially,
thus preventing immediate artificial fragmentation,
while the second one states that the initial density
fluctuations are too small to immediately trigger the formation of
sink particles.
As the system evolves, the density $\rho$ increases in some regions, 
and since $n_J\propto M_J\propto\rho^{-1/2}$ for an isothermal
gas, $n_J$ decreases in the same regions.
Eventually the conditions~(\ref{condaf}) and~(\ref{condc})
will both be violated
in dense regions. If condition~(\ref{condc})
is violated first, a sink particle will be created, and this will prevent
artificial fragmentation from happening. If instead the
condition~(\ref{condaf}) is
violated first, then artificial fragmentation will occur, but as long
as the fragments remain close to one another, they might eventually
get lumped together into a single sink particle once that sink is created.
The creation of a sink particle requires an increase in density by a factor
$f_{\rm sink}=\rho_c/\bar\rho$, while violating the Jeans criterion requires
an increase in density by a factor 
$f_{\rm a.f.}=(n_{J,\rm init}/n_{J,\rm a.f.})^2=(N/N_Jn_{J,\rm a.f.})^2$. 
For the benchmark simulation, we get $f_{\rm sink}=40,000$ and
$f_{\rm a.f.}=81$. Hence, the Jeans criterion will be violated
after an increase in density by a factor of 81, long before sink
particles are created. Artificial fragmentation must be prevalent in
the isothermal simulations of KB, but these fragments might get
re-aggregated when sink particles are created.\footnote{Actually, the
limited resolution of the gravity solver might help to prevent
artificial fragmentation in the simulations of KB (Klessen 2002).}.

There are four possible solutions to the problem of artificial fragmentation.
The most obvious solution (1) consists
of reducing $\rho_c$ down to $81\bar\rho$ or less, so that fragments would
turn into sink particles before they are dense enough to 
experience artificial fragmentation. The obvious drawback of this approach is
that the simulation would completely miss the late-stage evolution of dense 
fragments. If we keep the threshold density at a value $\rho_c=40,000\bar\rho$,
we must then find a way to increase the number of particles per Jeans mass.
We could (2) increase the initial Jeans mass $M_{J,{\rm init}}$
by a factor of 
$(40,000/81)^{1/2}\sim22$, so that by the time the density reaches $\rho_c$,
the Jeans mass would still contain enough particles to be properly resolved.
However, for a fixed initial number $N$ of particles, this would
reduce the initial number of Jeans masses in the
system from $N_J=222$ down to $N_J=10$, leading to poor statistics,
as very few cores would form.

If we keep both $\rho_c$ and $N_J$ fixed, we must then reduce the particle
mass by a factor of at least $(40,000/81)^{1/2}$ to insure sufficient
resolution. If (3) this is done over the entire computational volume, the
number of particles would then increase from 200,000 to 4,400,000. This
is clearly a brute-force approach, that would result in a tremendous increase
in computational time. The better approach (4) consists of reducing the 
particle mass {\it locally}, in dense regions where the Jeans mass is small.
This solution can be achieved dynamically, using particle splitting.

After splitting, the mass per particle is given by
\begin{equation}
\label{msplit}
m=m_{\rm init}f_{\rm split}^{-N_{\rm gen}}\,,
\end{equation}

\noindent where $m_{\rm init}=M_{\rm tot}/N$ is the initial mass of 
the particles, $f_{\rm split}$ is the splitting factor, and $N_{\rm gen}$
is the maximum number of ``generations,'' that is, the maximum number of
splittings a particle can experience. At the threshold density $\rho=\rho_c$,
the Jeans mass is given by
\begin{equation}
\label{mjc}
(M_J)_c=M_{J,{\rm init}}\left({\bar\rho\over\rho_c}\right)^{1/2}
={Nm_{\rm init}\over N_J}\left({\bar\rho\over\rho_c}\right)^{1/2}\,.
\end{equation}

\noindent We eliminate $m_{\rm init}$ in equation~(\ref{mjc}) using
equation~(\ref{msplit}). The Jeans criterion, 
$(M_J)_c/m\geq n_{J,{\rm a.f.}}$ becomes
\begin{equation}
\label{njmax}
{N\over N_J}\biggl({\bar\rho\over\rho_c}\biggr)^{1/2}
f_{\rm split}^{N_{\rm gen}}\geq n_{J,\rm a.f.}\,.
\end{equation}

\noindent This relates the initial number of Jeans masses $N_J$ to
the number of splitting generations $N_{\rm gen}$ necessary to
prevent artificial fragmentation. This expression is useful 
for generating initial conditions (see \S4.2 below).

\subsection{Scenarios}

\begin{figure}[t]
\vskip-0.4in
\hskip0.3in
\includegraphics[width=5.8in]{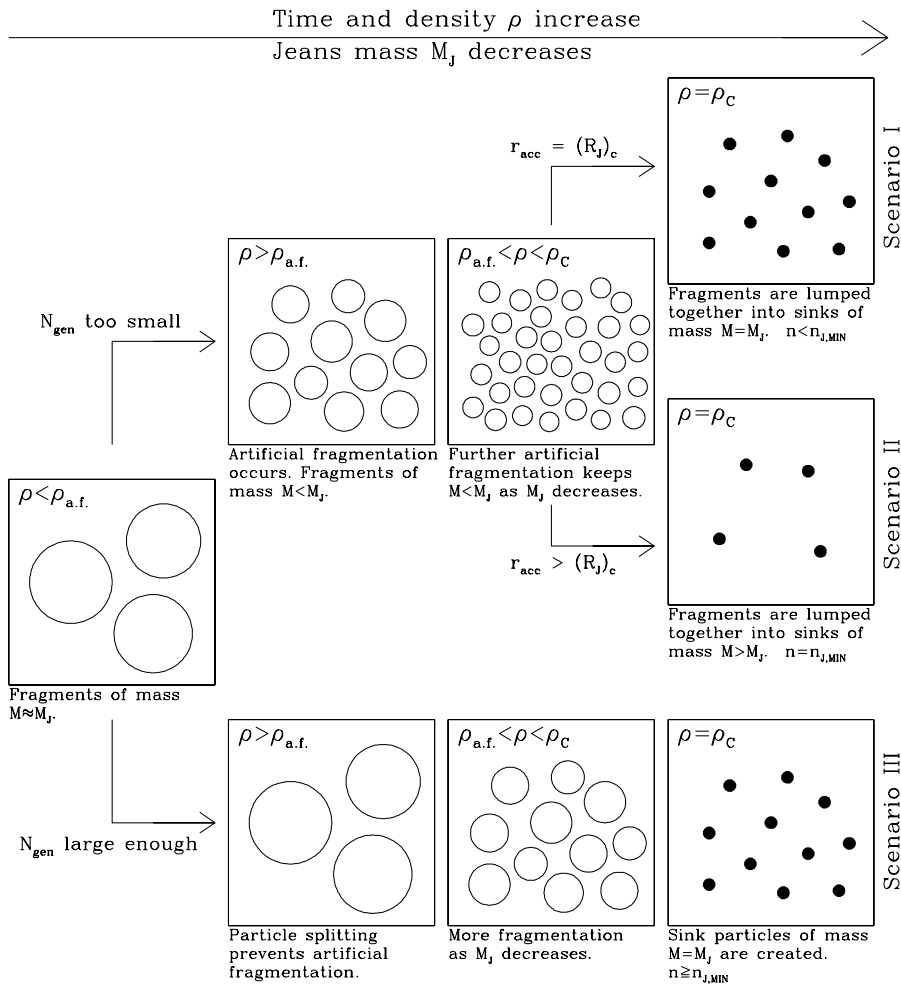}
\vskip-0.4in
\caption{Schematic illustration of the various scenarios. Open circles
represent fragments composed of SPH particles. Solid dots represent
sink particles. {From} left to right, time increases,
local density increases, and Jeans mass decreases.}
\end{figure}

The evolution and fragmentation of the cloud can follow three different
paths, which we call {\it scenarios}. The particular scenario followed
is determined by two parameters: the maximum number of splitting generations
$N_{\rm gen}$ and the accretion radius $r_{\rm acc}$, which determines,
among other things, the number of gas particles that are lumped together
when a sink is created.
Figure~1 illustrates the three different scenarios. 
Consider first a simulation with an insufficient number
of splitting generations, or no splitting at all ($N_{\rm gen}$ too small).
As the system evolves, the density increases and the Jeans mass decreases
in overdense regions. Eventually, the
cloud fragments into Jeans-mass clumps, which
themselves fragment as the density increases and $M_J$ decreases,
until the density reaches $\rho_{\rm a.f.}\equiv f_{\rm a.f.}\bar\rho$.
At that point, artificial fragmentation occurs, 
leading to fragments of mass $M<M_J$.
As $\rho$ increases and $M_J$ decreases, the mass of
these ``artificial fragments''
might eventually end up above $M_J$, but further artificial
fragmentation will occur, bringing the mass per fragment below
$M_J$ again. Finally, when the density reaches the threshold
value $\rho_c$, sink
particles are created. KB suggest setting the accretion radius $r_{\rm acc}$
equal to the Jeans length $(R_J)_c$ at the density $\rho=\rho_c$.
Each sink particle will then contain about one
Jeans mass, implying that several sub-Jeans-mass
fragments will be lumped together into the same sink particle, possibly
nullifying the effects of artificial fragmentation. This is Scenario~I.

One concern with this scenario is that the Jeans mass $(M_J)_c$ at
the threshold density $\rho_c$ 
might be greatly underresolved (whether or not
artificial fragmentation occurred), and replacing it
by a sink particle might be inappropriate.
For instance, the benchmark simulation of KB contains 200,000 particles
and starts up with 222 Jeans masses, or $n_{J,\rm init}=901$ particles
per Jeans mass. The creation of sink particles requires an increase in
density by a factor of 40,000, corresponding to a
decrease in Jeans mass by a factor $40,000^{1/2}=200$. Hence, by the
time sinks are created, the Jeans mass is down to $901/200\sim5$ particles,
which is clearly insufficient to resolve it. 

Scenario II also describes simulations without particle splitting, or with
an insufficient number of splitting generations, but differs
from Scenario~I in the
choice of the accretion radius $r_{\rm acc}$. In this scenario,
$r_{\rm acc}$ is set to
a value larger than $(R_J)_c$, by requiring that
each sink particle must be made of at least $n_{J,\min}$ particles, as
described in \S2.4. As a result, 
fewer sink particles are created, and their initial masses
(that is, before they grow by accretion) exceed 
the Jeans mass by a factor $n_{J,\min}/(n_J)_c$. 

Finally, Scenario III describes simulations with a sufficient
number of splitting generations.
Particle splitting prevents artificial fragmentation, and when sink
particles are created, each Jeans-mass fragment contains enough
particles to be replaced by a sink particle, without lumping
fragments together. The number of sink particles formed
under Scenarios I and III should be comparable, since each sink
particle is created with a mass $M\sim (M_J)_c$. However, in Scenario III,
the Jeans mass was fully resolved throughout the entire
calculation, thus preventing artificial fragmentation, while in Scenario I,
artificial fragmentation leads to sub-Jeans-mass fragments, which 
{\it presumably} get lumped together when sinks are created.

Notice that Scenario~III is the only one we regard as satisfactory. Scenarios~I
and II suffer from artificial fragmentation. Also, in Scenario I underresolved
fragments are turned into sinks, while in Scenario~II Jeans-mass objects
($M\gtrsim M_J$) are not allowed to form.

\section{THE SIMULATIONS}

\subsection{Initial Conditions}

Our method for generating initial conditions is similar to the one used
by KB, and is based on the Zel'dovich approximation commonly used
for cosmological simulations. We assume that the
initial density is described by a Gaussian random field with
a density power spectrum $P(k)$, where $k$ is the wavenumber.
The details are given in Appendix A. As in KB, we initially
consider a power spectrum
that follows a power law, $P(k)\propto k^{-n}$, with $n=2$.
The rms density fluctuation $\delta_\lambda$ at scale $\lambda\sim1/k$
is given by
$\delta_\lambda\sim k^{3/2}P^{1/2}(k)\sim k^{1/2}\sim\lambda^{-1/2}$.
Hence, the density fluctuations are larger at smaller scales.

\subsection{Numerical Parameters}

\begin{deluxetable}{ccccccccc}
\tabletypesize{\footnotesize}
\tablecaption{Numerical Parameters of the Calculations\label{tbl-1}}
\tablewidth{0pt}
\tablehead{
\colhead{Run} & \colhead{$N_{\rm gen}$} & \colhead{$(n_J)_c$} & 
\colhead{$n_{J,\min}$} & \colhead{$n_{\rm acc}$} &
\colhead{$(N_J)_{\rm sink}$} & \colhead{$r_{\rm acc}/L_{\rm box}$} & 
\colhead{$r_{\min}/\Delta x$} & \colhead{Scenario}
}
\startdata
A & 0 &   3 & 100 & 100 & 38.1 & 0.00132 & 0.30 &  II \\
B & 1 &  21 & 100 & 100 &  4.8 & 0.00066 & 0.16 &  II \\
C & 2 & 168 & 100 & 168 &  1.0 & 0.00039 & 0.10 & III \\
\enddata
\end{deluxetable}

For all simulations presented in this paper, we use $N=64^3=262,144$,
$n_{J,\min}=n_{J,{\rm a.f.}}=100$, 
$\rho_c/\bar\rho=40,000$, and $f_{\rm split}=8$.
We allow up to 2 generations of particle splitting.
With $N_{\rm gen}=2$, equation~(\ref{njmax}) gives 
$N_J=839$ to be the maximum
number of Jeans masses the system can contain initially. This is a rather 
large number, and it might be desirable to use a smaller one to 
increase the resolution per fragment, while retaining good statistics.
We shall use instead $N_J=500$ in all simulations. The initial number
of particles per Jeans mass is then $n_{J,{\rm init}}=N/N_J=524$.
We performed three simulations, Runs A, B, and C, with identical initial
conditions and $N_{\rm gen}= 0$, 1, and 2, respectively, as indicated in 
Table~1. All other input parameters, including $\rho_c/\bar\rho$, are the same
for all runs, but
the accretion radius $r_{\rm acc}$ is adjusted such that the sinks,
at the time of their creation, contain at least 1 Jeans mass, and at
least enough particles to be properly resolved
(that is, $n_{\rm acc}=\max[(n_J)_c,n_{J,\min}]$). With this particular
choice, Runs~A and B, which do not have enough generations of particle 
splitting, will follow Scenario~II; Run~C, with ($N_{\rm gen}=2$),
will follow Scenario~III.

\begin{figure}[t]
\vskip-1.4in\hskip0.1in
\includegraphics[width=6.4in]{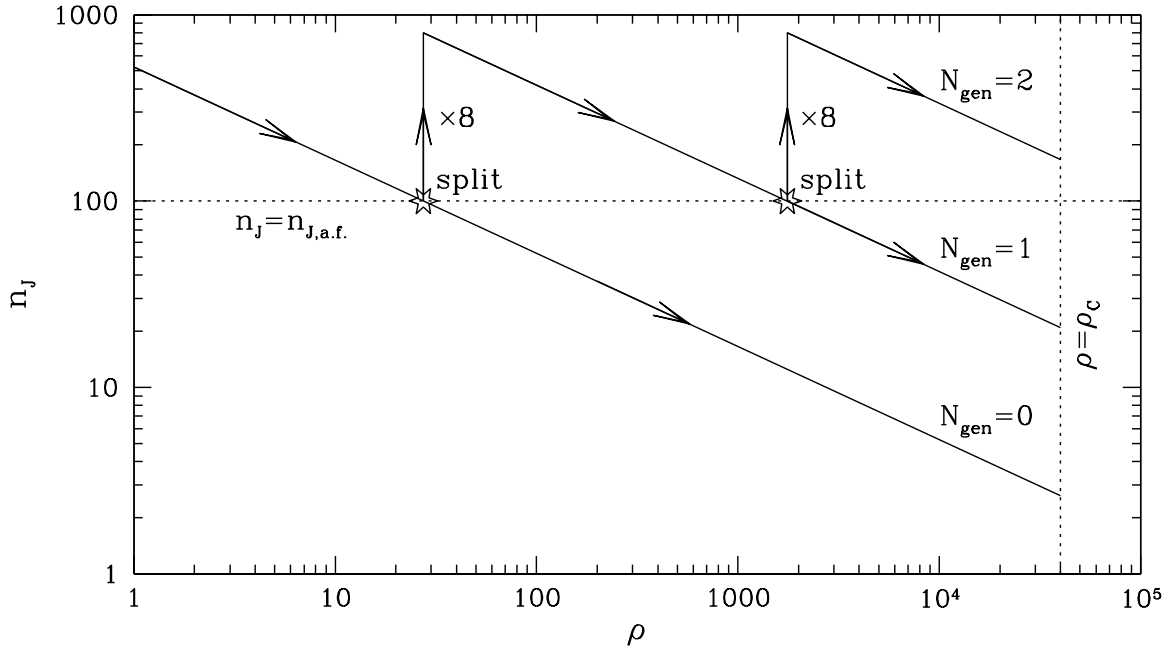}
\vskip-1.5in
\caption{Evolution of the number of particle per Jeans mass,
$n_J$, as the density $\rho$ increases, for three different values
of $N_{\rm gen}$, 0 (no splitting), 1, and 2, as indicated.
The horizontal dotted line indicates $n_{J,\rm{a.f.}}$, the
minimum value of $n_J$ required to prevent artificial fragmentation.
Particle splitting, when allowed, occurs when $n_J$ drops below
$n_{J,\rm{a.f.}}$, and causes $n_J$ to increase by a factor of
$f_{\rm split}=8$. The vertical dotted line indicates the threshold
density $\rho=\rho_c$ for creating sink particles. The solid dots
on that line indicate the number of particles that end up inside
each sink particle at the time of its creation, according
to the various Scenarios (I, II, and III).}
\end{figure}

Figure 2 shows the evolution of the number $n_J$ of
particles per Jeans mass as the density increases.
This number decreases as $\rho^{-1/2}$, and drops below
$n_{J,{\rm a.f.}}=100$ when the density reaches $\rho=(524/100)^2=27.5$.
Without particle splitting ($N_{\rm gen}=0$), $n_J$ will drop down to
$(n_J)_c\sim3$
when the density reaches $\rho=\rho_c$. 
The accretion radius $r_{\rm acc}$ is set to
a value larger that $(R_J)_c$, such that groups of
$n_{J,\min}=100$ particles will be converted into sink particles of
mass $M_{\rm sink}\approx38M_J$. With $N_{\rm gen}=1$, particles will
split once their density reaches $\rho=27.5$ (first star in Fig.~2),
increasing $n_J$ from 100
to 800. Then, $n_J$ will keep dropping as $\rho$ increases, reaching
$(n_J)_c\sim21$ at $\rho=\rho_c$. Groups of 100 particles will then be
converted into sink particles of mass $M_{\rm sink}\approx4.8M_J$,
Finally, with $N_{\rm gen}=2$, particles
will split a second time when the density reaches $\rho=1759$
(second star in Fig.~2),
increasing $n_J$ to 800 again. Eventually, $n_J$ will drop to $(n_J)_c=168$
when the density reaches $\rho=\rho_c$ and groups of 168 
particles will be
converted into sink particles of mass $M_{\rm sink}\approx M_J$.
Hence, in a simulation with a sufficient number of splitting generations,
the number of particles $n_J$ in a Jeans mass follows a seesaw pattern,
always staying above $n_{J,{\rm a.f}}$ and thus preventing artificial
fragmentation; Jeans masses are always properly resolved.

The accretion radius $r_{\rm acc}$ must be determined experimentally.
We ran the code with various values of $r_{\rm acc}$, up to the
point when a few sinks (20 or so) have formed. We then check how
many gas particles were removed when each sink was created.
There is an ``optimal'' number of particles, which is the maximum
of $(n_J)_c$ and $n_{J,\min}$ (100 for Runs A and B;
168 for Run C). If the number of particles exceeded
significantly that optimal number, this indicated that $r_{\rm acc}$
was too large. We would then try with a smaller value, and iterate
until the number of particles turned into each sink was close to
the optimal number. Notice that it could not be smaller, because the
code would then delay the formation of sinks until enough particles have
fallen inside $r_{\rm acc}$.

Once $r_{\rm acc}$ is determined, we must ensure that the resolution of
the algorithm is sufficient to resolve that length scale. For the
hydrodynamics, this is achieved by allowing the smoothing length of the
SPH particles to shrink down to values smaller than $r_{\rm acc}$ in
dense regions. For the gravity, the particle-mesh part of the $\rm P^3M$
algorithm uses a $128^3$ grid to calculate the gravitational force. 
The corresponding length resolution is about
$2\Delta x\approx0.016L_{\rm box}$, where $\Delta x=L_{\rm box}/128$ 
is the cell
size. The short-range correction part of the algorithm extends the resolution
below the cell size, and the softening length $r_{\min}$ can be chosen
arbitrarily. We choose $r_{\min}$ to be slightly smaller than $r_{\rm acc}$.
With these particular choices, the gravitational force will be properly
resolved at all scales down to the scale 
corresponding to sink formation. The hydrodynamical forces
will also be properly resolved, as long as the smoothing lengths shrink
down to a value comparable to $r_{\min}$ or less. This will happen only
if the mass resolution is large enough, that is, a Jeans mass contains
at least $n_{J,{\rm a.f.}}$. If this is not the case, however,
the hydrodynamical forces will be underresolved, and {\it this is
precisely what causes artificial fragmentation}.

In low density regions where
the smoothing lengths $h$ are much larger, there is clearly a
mismatch between the resolutions of the gravity and the pressure force.
This should not matter much, since particles are widely
separated in these regions, and both gravitational and pressure
forces are properly resolved at that scale.
We have not observed any artificial clumping
of SPH particles in low density regions resulting from the gravity
being overresolved compared to the hydrodynamics (see Fig.~5 below).

Table~1 lists the values of
$(n_J)_c$, $n_{J,{\min}}$, $n_{\rm acc}$, $r_{\rm acc}$, $r_{\min}$,
and the number 
$(N_J)_{\rm sink}=n_{\rm acc}/(n_J)_c$ of Jeans masses
inside sink particles at the time of their creation. Notice that the
simulation with $N_{\rm gen}=2$ is the only one that forms
Jeans-mass objects [$(N_J)_{\rm sink}=1$].

\subsection{Computational and Physical Units}

The calculations are performed in computational units. 
In these units, the total mass $M_{\rm tot}$ of the
system, the box size $L_{\rm box}$, and the gravitational
constant $G$ are unity. Hence, the mean density $\bar\rho$ is
also unity, and time is expressed in units of
$(G\bar\rho)^{-1/2}$.
The initial Jeans mass is given by
\begin{equation}
\label{mjinit}
M_{J,{\rm init}}=\biggl({5kT\over2G\mu}\biggr)^{3/2}
\biggl({4\pi\bar\rho\over3}\biggr)^{-1/2}
\end{equation}

\noindent where $k$ is the Boltzman constant, and $\mu$ is the mean
mass per molecule (Tohline 1982). In computational units, this reduces to
\begin{equation}
\label{mjinit2}
M_{J,{\rm init}}=\biggl({5\epsilon\over3}\biggr)^{3/2}
\biggl({3\over4\pi}\biggr)^{1/2}=1.0513\epsilon^{3/2}\,,
\end{equation}

\noindent where $\epsilon=3kT/2\mu$ for a gas with $\gamma=5/3$.
Since $M_{J,{\rm init}}=M_{\rm tot}/N_J=1/N_J$, this reduces to
\begin{equation}
\label{epsinit}
\epsilon=0.9672N_J^{-2/3}\,.
\end{equation}

\noindent This is the prescription for choosing the initial
value of $\epsilon$, which
then remains constant throughout the calculation under the
assumption of isothermality.
With our particular choice of $N_J=500$, we get $\epsilon=0.01535$.

The Jeans criterion is given by
\begin{equation}
{M_J\over m}\geq n_{J,{\rm a.f.}}\,.
\end{equation}

\noindent We set $M_J=M_{J,{\rm init}}(\rho/\bar\rho)^{-1/2}
=M_{J,{\rm init}}/\rho^{1/2}$, and eliminate $M_{J,{\rm init}}$
using equation~(\ref{mjinit2}). We get
\begin{equation}
\label{jccomp}
{\epsilon^3\over\rho m^2}\geq{36\pi\over125}n_{J,{\rm a.f.}}^2\,.
\end{equation}

\noindent Whenever the internal energy, density, and mass of a particle
violate this condition, that particle is split by the algorithm
(unless this would exceed the maximum number of splitting generations
$N_{\rm gen}$ allowed).

The simulations are scale-free, and
can be rescaled to any physical size of interest. To rescale to
physical units, we choose particular values for the temperature $T$
and mean density $\bar\rho$ of the cloud, and compute the initial
Jeans mass $M_{J,{\rm init}}$ using equation~(\ref{mjinit}).
The total mass of the
system is then $M_{\rm tot}=N_JM_{J,{\rm init}}$, where $N_J$ was the
value used in equation~(\ref{epsinit}) to set the initial conditions,
the size of the box is $L_{\rm box}=(M_{\rm tot}/\bar\rho)^{1/3}$, and
the physical time is obtained by multiplying the computational
time by $(G\bar\rho)^{-1/2}$. This will be illustrated in \S7 below, 
where we rescale the results of our simulations to several physical
densities and compare to particular physical systems of interest.

\section{RESULTS}

Notation in this field is not standardized. For this paper, we adopt
the following terminology, partly to facilitate comparison to the work of
KB. A ``dense region'' is a part of a molecular cloud that is likely to
form stars; for this paper, we focus on massive, dense regions able to
form clusters. A ``clump'' is a region of enhanced density in the original
dense region. A ``core'' is a clump that has become gravitationally unstable.
We will later identify cores with sink particles, and we will interpret
the mass function of cores in terms of the mass function of stars and
substellar objects. Note that observers often use the work ``core'' to
describe what we call here ``dense regions.''

\subsection{Global Evolution}

\begin{figure}
\hskip.8in
\includegraphics[width=5in]{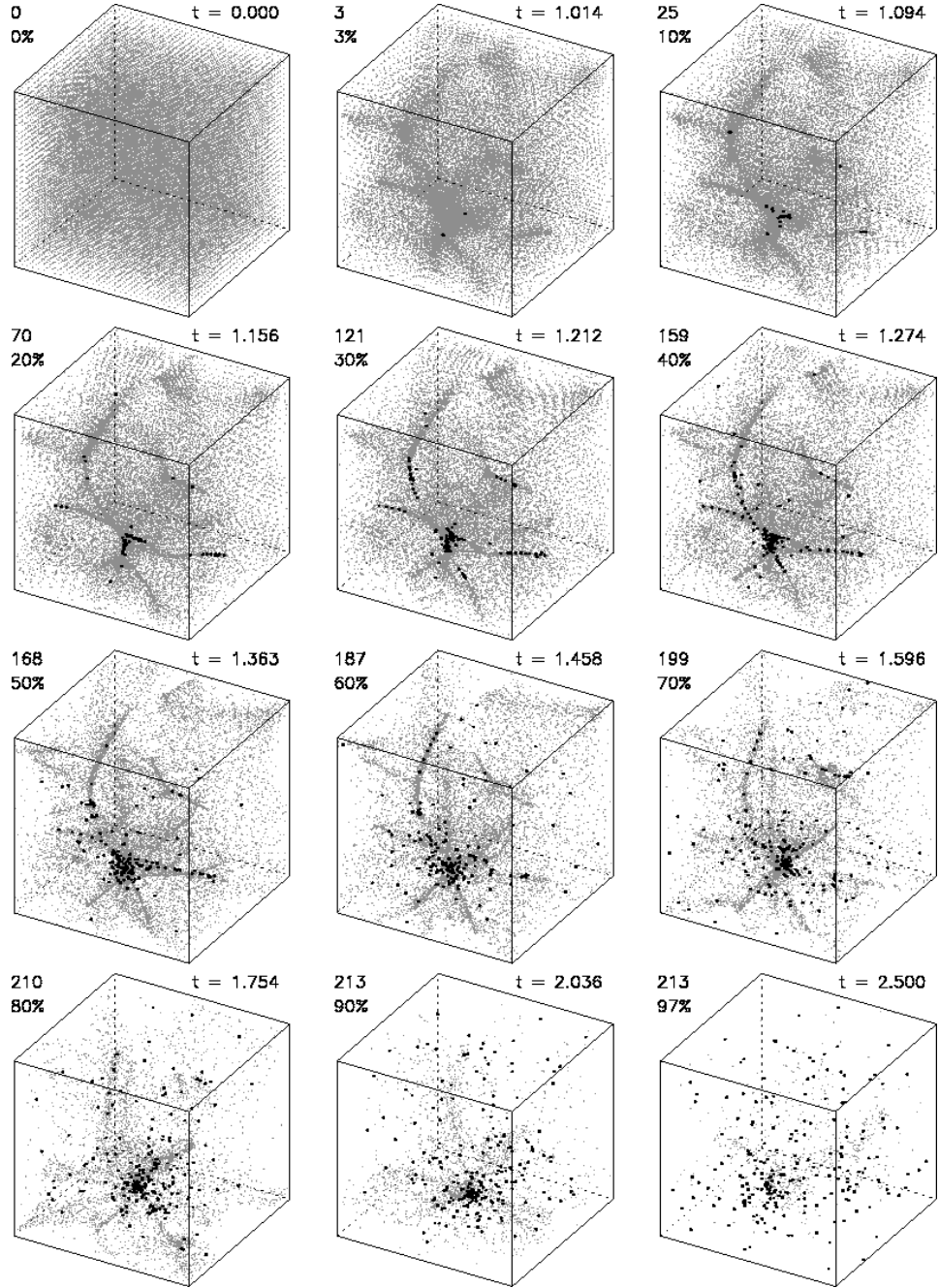}
\caption{Global evolution of the system, for Run A. Each box shows
a snapshot of the system, with SPH gas particles represented by blue dots
(for clarity, only 1/8 of the gas particles are plotted), and 
protostellar cores represented by large black dots. For each snapshot, the time
in units of $(G\bar\rho)^{-1/2}$ is indicated in the top right corner,
while the numbers in the top left corner indicate the number of cores
and the fraction of the total gas that has been accumulated into cores,
respectively.}
\end{figure}

\begin{figure}[t]
\hskip.8in
\includegraphics[width=5in]{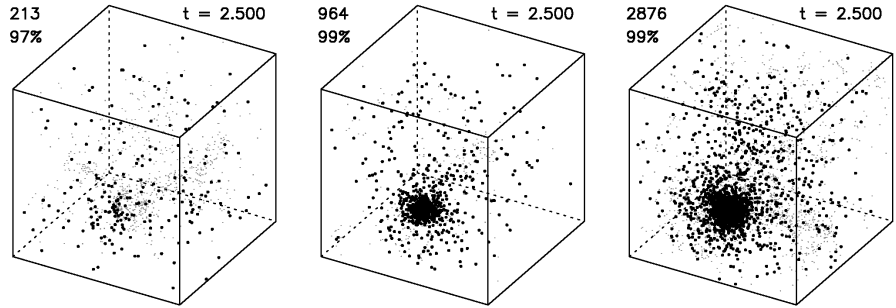}
\vskip0.0in
\caption{Final state of the system, for Run A (left), B (middle), and
C (right). The symbols and labels have the same meaning as in Figure~3.}
\end{figure}

\begin{figure}[t]
\hskip1.1in
\includegraphics[width=4.25in]{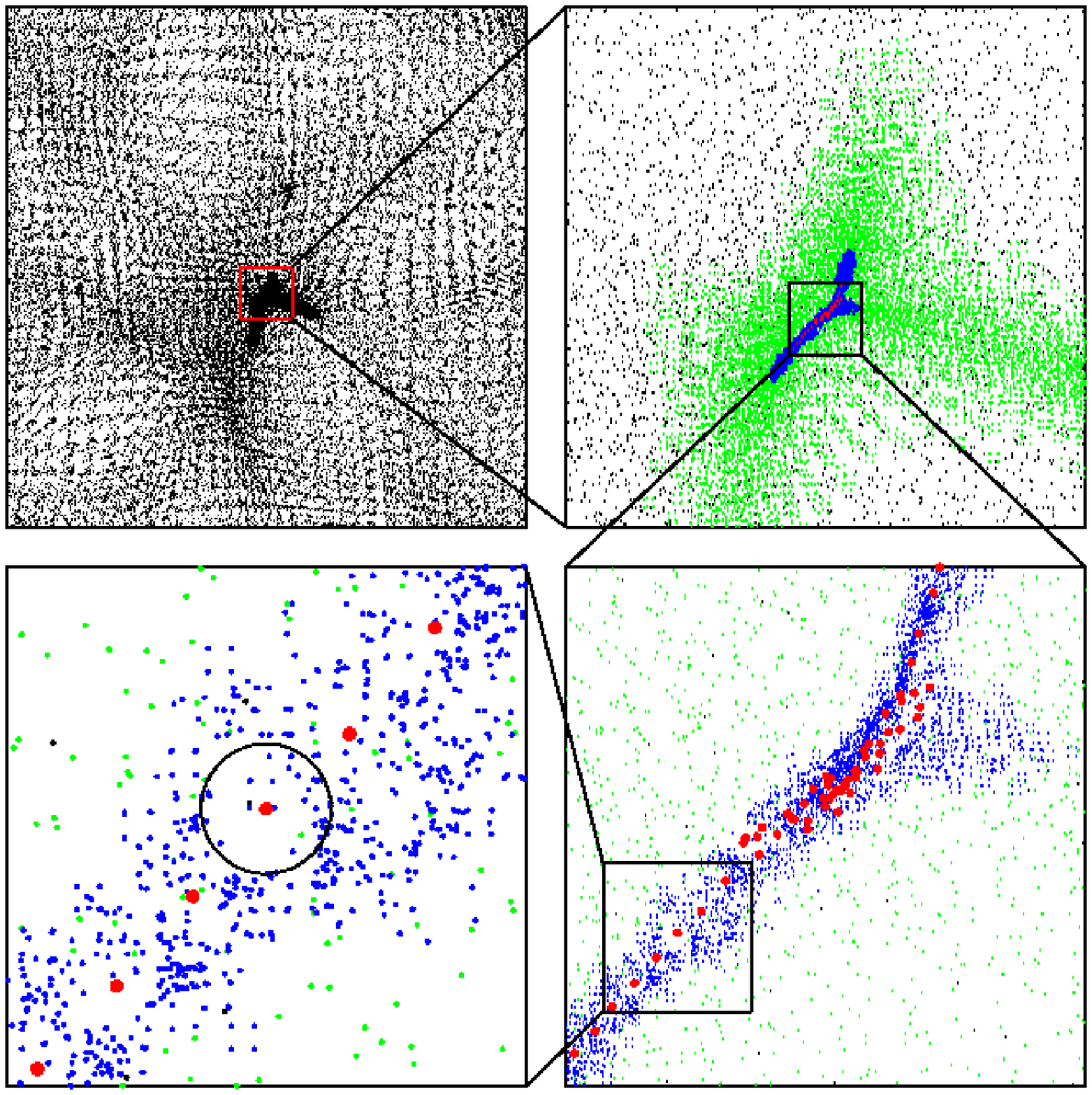}
\vskip0.1in
\caption{Early time slice ($t=0.903$)
of the evolution of the system, for Run C.
Top left panel: the entire computational volume. For clarity, only 1/8
of the particles are shown. Top right panel: zoom-in on a dense region at
the intersection of two emerging filaments (with all particles shown).
The colors black, green, and blue corresponds to particles of
generation 0, 1, and 2, respectively. Sink particles are shown in red.
Bottom right panel: zoom-in on the central, high-density region. Cavities
around sink particles are visible. Bottom left panel: Zoom-in on a filament.
The cavity around one sink particle is shown by a circle of radius
$r_{\rm acc}=0.0005$. Particles appearing inside the cavity are either
background or foreground particles.}
\end{figure}

Figure 3 shows the evolution of the system for Run A ($N_{\rm gen}=0$). 
Gas particles are
shown in blue, and sink particles in black. 
Following KB, we shall identify these sinks as {\it protostellar cores},
and from now on we shall use the term ``sink'' only when discussing
the properties of the algorithm, rather than the physical 
interpretation of the results. Each panel is accompanied by
three numbers, the time, the number of cores, and the percentage
of mass accumulated in cores. The system evolves rapidly into a network
of intersecting filaments. Around $t=1$, cores start to form inside the
dense knots located at the intersections of the filaments. By $t=1.2$, the
densest knot already contains a large cluster of cores, and the filaments
themselves start to fragment into cores. By $t=1.6$, most of the filaments
have disappeared, having turned into cores, and these cores are
falling toward the main cluster. By $t=2$, 90\% of the gas has been turned
into cores, and a dense cluster of cores has formed. During the later stages
of the evolution, the remaining gas gets accreted onto the cores, and
the cluster
starts to evaporate: cores are ejected by close encounters, while the
reminder of the cluster contracts and gets more tightly bound. All
calculations terminate at $t=2.5$, when 97\%--99\% of the
gas has been turned into cores.

Figure 4 shows the final state of the system, at $t=2.5$, for all three runs.
The final location of the cluster is about the same for all three runs, but
the total number of cores significantly increases as we allow more splitting
generations. The final number of cores is 213, 964, and 2876 for Runs A, B,
and C, respectively. Also, there is 
less gas remaining in the system for Runs B and C, compared to Run A.

Figure 5 shows a series of zooms at an early time slice ($t=0.9030$)
for Run~C ($N_{\rm gen}=2$). The black, green, and blue dots represent
the generation~0 (unsplit), generation~1 (split once), and generation~2
(split twice) particles. The masses of these different types of particles
have ratios (64:8:1). As explained in \S4.2 above, the first splitting
occurs at density $\rho=27.5$ and the second one at density
$\rho=1759$. Hence, the boundary between the black and green particles is
an isosurface of density $\rho=27.5$, and the boundary between 
green and blue particles is an isosurface of density $\rho=1759$. Notice
that these boundaries are quite sharply delimited, indicating that
nothing peculiar is happening there. Indeed, the density varies smoothly
across this entire region, and the transition between particles of
widely different masses produces no observable
feature in the density profile.

The bottom right panel of Figure~5 zooms in on the intersection of
filaments to show a concentration of sinks (red circles), which have 
formed earlier and have moved toward each other as the density increases. 
In this region, a
gas particle might be located inside the accretion radius of several
sinks, in which case it will accrete onto the sink for which the total
energy is the lowest (that is, the binding energy is highest), 
as explained in \S2.4. The bottom left panel shows a further zoom-in
on the filament to the lower left.  In this particular case,
the cylindrical collapse of a filament leads to the formation of
a chain of sink particles separated by a distance
equal to twice the accretion radius $r_{\rm acc}$, reminiscent of
observations of dense regions within filaments \citep{schneider79}.
The circle shows the accretion radius for a particular sink. Gas
particles that enter that accretion radius will be accreted by the sink
(most particles appearing inside the circle in Fig.~5 are seen in projection).

\subsection{Particle Splitting and Sink Particles}

\begin{figure}[t]
\vskip-0.4in
\hskip-0.in
\includegraphics[width=6.2in]{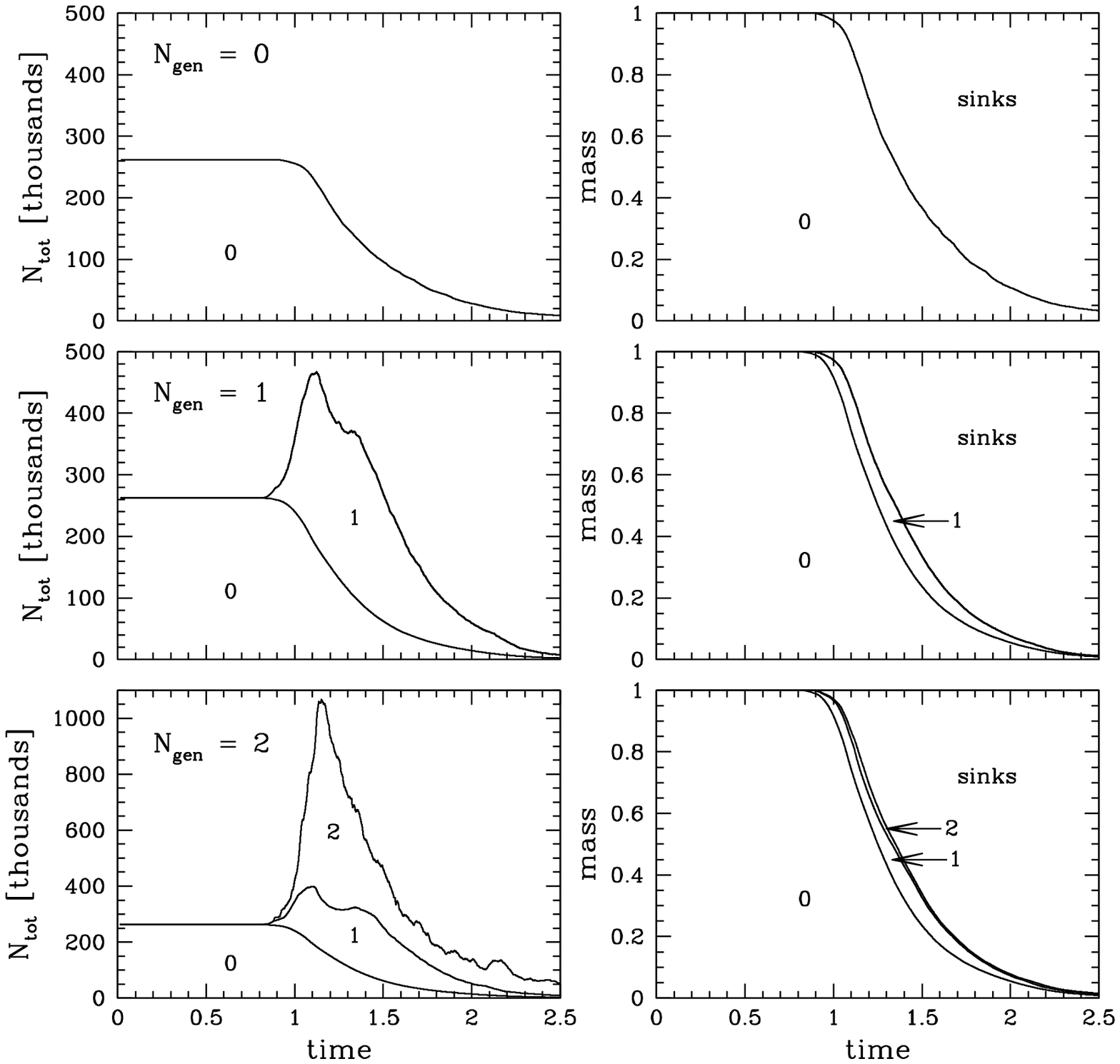}
\vskip-0.8in
\caption{Left panels: stacked histograms of the number of particles of
various generations, versus time. The labels 0, 1, and 2 indicate
the various generations of particles. The top curve on each
panel shows the total number of particles, $N_{\rm tot}$.
Right panels: stacked histograms of the mass contained in each
generation of particles, and in sinks, versus time.
Top panels: Run A; middle panels: Run B; bottom panels: Run C.}
\end{figure}

\begin{deluxetable}{crrrc}
\tabletypesize{\footnotesize}
\tablecaption{Number of Particles.\label{tbl-2}}
\tablewidth{0pt}
\tablehead{
\colhead{Run} & \colhead{$N_{\rm init}$} & \colhead{$N_{\rm eff}$} & 
\colhead{$N_{\rm peak}$} & \colhead{$N_{\rm peak}/N_{\rm eff}$}
}
\startdata
A & 262,144 &    262,144 &   262,144 & 1.000 \\
B & 262,144 &  2,097,152 &   467,636 & 0.223 \\
C & 262,144 & 16,777,216 & 1,067,610 & 0.064 \\
\enddata
\end{deluxetable}

Figure~6 shows stacked histograms of the number of particles within
each generation (left panels) and the
total mass contained within each generation, and the division of mass
between particles and sink particles (right panels). In the
absence of particle splitting ($N_{\rm gen}=0$, Run A, top panels),
the number of particles remains constant until sink particles are
created, and then steadily decreases as sink formation and
accretion onto sinks
removes particles from the simulation. When particle splitting is
included (Runs B and C, middle and bottom panels),
the total number of particles $N_{\rm tot}$ initially increases as
particles split, but this process competes with accretion onto sinks,
and eventually $N_{\rm tot}$ decreases.
In Table~2, we list, for each run, the initial number of particles
$N_{\rm init}$, the maximum number of particles $N_{\rm peak}$
reached during the simulation, and the effective number of particles
$N_{\rm eff}$, defined as the number of particles a simulation without
particle splitting would need to have the same resolution
($N_{\rm eff}=N_{\rm init}f_{\rm split}^{N_{\rm gen}}$). Runs A, B, and C have
effective resolutions of $64^3$, $128^3$, and $256^3$ particles,
respectively. Of particular interest is the ratio $N_{\rm peak}/N_{\rm eff}$,
given in the last column of Table~2. This ratio measures the ``efficiency''
of the particle splitting algorithm. As the number of splitting generations
increases, the peak number of particles becomes significantly lower than
the effective number of particles, resulting in a substantial saving
of computational effort relative to a simulation without particle splitting.
{From} Table~2, we can infer that every additional splitting generation makes
the algorithm more ``efficient'' by a factor of order $4-5$.

At this point, we need to discuss a possible alternative to particle
splitting, called the ``zoom-in'' approach, which is often used in
numerical simulations. This approach would consist of first running
a low-resolution, $N=64^3$ simulation, identifying in the final
state of the simulation the particles located in
``regions of interest,'' where higher resolution would be desirable,
going back to the initial conditions, replacing {\it these particles
only} by a set of smaller particles, and then redoing the simulation.
This approach would fail in the present case for the following reason:
As we shall see in the next section, about half of the gas particles in
the simulation are eventually converted into sinks, while the other half
get accreted onto existing
sinks. While high resolution might not be necessary
for the particles that are accreted onto sinks, it is certainly
necessary for the particles that are converted into sinks, in order to
prevent artificial fragmentation. This means that, in the initial conditions,
one half of the particles would have to be replaced by a cube of 
$4\times4\times4=64$ particles in order to provide sufficient resolution.
The total number of particles would then be $N=64^3[1/2+(4^3-1)/2]=8,388,608$,
or $256^3/2$. Hence, using a zoom-in approach would only reduce the
number of particles by a factor of 2 compared with the brute force
approach. The zoom-in approach is useful in situations where the ``regions
of interest'' contain a small fraction of the total mass of the system.
Here, these regions contain 50\% of the mass, making this approach
inefficient.

\subsection{Regimes}

\begin{figure}[t]
\vskip-.4in
\hskip0.2in
\includegraphics[width=5.7in]{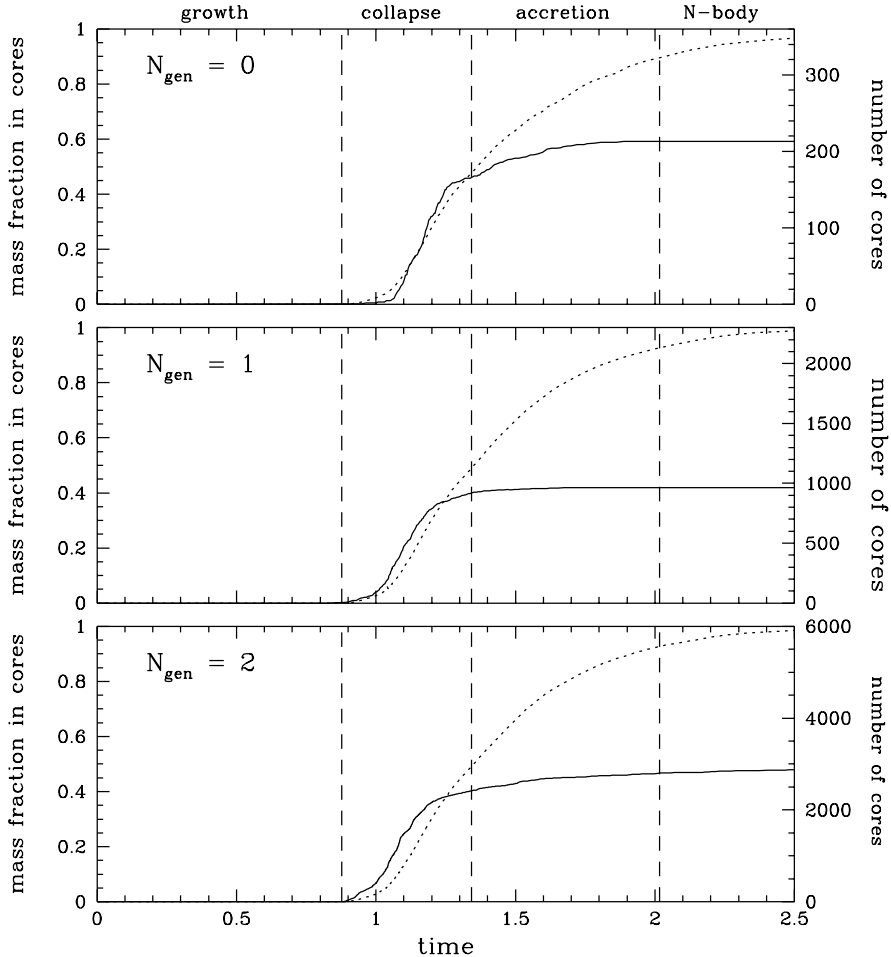}
\vskip-0.3in
\caption{Time-evolution of the mass accumulated in protostellar
cores (dashed curve,
left axis) and the number of cores (solid curve, right axis) versus time.
The dotted lines separate the various phases of the evolution, with the
corresponding regimes labelled on top.}
\end{figure}

Figure 7 shows the time-evolution of the number of cores and the total mass
accumulated
in cores. By comparing the time-evolution of these two quantities,
we can identify four distinct phases during the evolution of the
cloud, with each phase corresponding to a different dynamical regime.

The initial phase corresponds to the {\it growth regime}. The initial
density fluctuations grow by gravitational instability, to form a network
of dense, intersecting filaments. This phase of the evolution terminates
when the first cores form. Since the density threshold $\rho_c$ for
sink formation is chosen arbitrarily, the end of this initial phase is
also arbitrary. However, in the absence of sinks (and with infinite
resolution), fragments would collapse and reach infinite densities at a
finite time that would exceed the time of sink formation only slightly.
Hence, choosing the time of formation of the first cores as corresponding to
the end of the growth phase is not an unreasonable choice.

After the first cores form, the evolution of the cloud enters the
second phase, which corresponds to the {\it collapse regime}. To
understand this regime, and the following one, we must consider a fundamental
property of Gaussian initial conditions: the filling factors of underdense
($\rho<\bar\rho$) and overdense ($\rho>\bar\rho$) regions are initially equal,
both being 1/2 of the total computational volume. Since the initial
density is nearly uniform ($\rho\approx\bar\rho$), 
about 1/2 of the gas starts up in overdense
regions, while the other 1/2 starts up in underdense regions. In the
collapse regime, the gas that started up in overdense regions collapses and
is converted into cores. This regime is characterized by an increase in both
the number of cores and the mass inside cores, at rates that are
roughly proportional.
This phase terminates when 50\% of the gas, essentially the gas that started
up in overdense regions, has been turned into cores.

The evolution of the cloud then enters the next phase, which corresponds to
the {\it accretion regime}. The uncollapsed gas remaining in the cloud is 
the gas that started up in underdense regions. The main
tendency of this gas is not to collapse onto itself
and form new cores, but rather to accrete onto the existing cores.
As a result, the mass in cores keeps increasing, while the number of cores
levels off. During this phase, the mass in cores nearly doubles, while
the number of sinks increases by less than 30\%.

Finally, once most of the gas (90\% or so) has been accumulated in cores,
the evolution of the cloud enters the fourth and last phase, which
corresponds to the {\sl N-body regime}. The hydrodynamics becomes irrelevant,
and the evolution of the system is governed by gravitational many-body
dynamics.

\subsection{Formation and Growth of Protostellar Cores; Low-Resolution 
Simulation}

\begin{figure}[t]
\vskip-0.1in
\hskip1.0in
\includegraphics[width=4.0in]{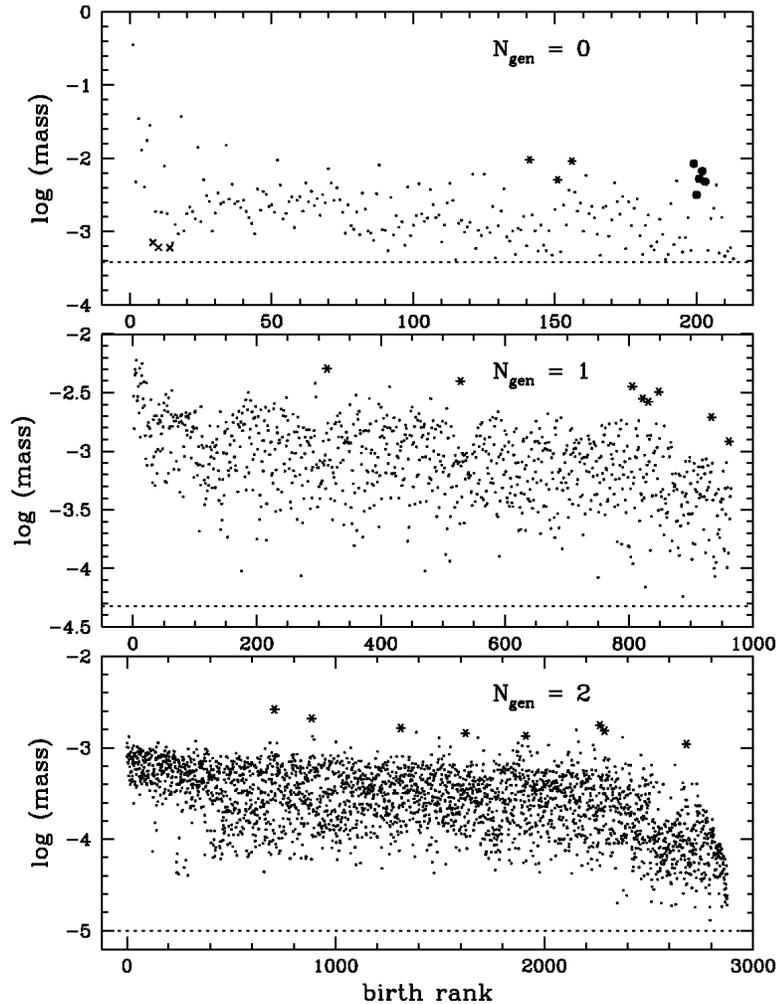}
\caption{Mass of protostellar cores versus birthrank. 
(a, top panel) Run A ($N_{\rm gen}=0$);
(b, middle panel) Run B ($N_{\rm gen}=1$);
(c, bottom panel) Run C ($N_{\rm gen}=2$).
The various symbols (crosses, asterisks, and solid circles) identify
particular cores that are discussed in the text.
}
\end{figure}

Figure 8 shows the final masses of the protostellar cores as a
function of their birth rank. We define the birth rank such that
the $n^{\rm th}$ core formed in the simulation has a birth rank
of $n$ (KB use the term ``index''). The top panel shows the results of 
Run A, without particle splitting, which can be directly compared to
the results of KB. As KB noted, the mass tends to decrease with
increasing birth rank, simply because cores formed later have less
time to grow by accretion. However, the trend we observe is much
less pronounced than the one found by KB. The mass range is comparable,
and the most massive cores all formed early (with the most massive
one being the very first one that formed), but the distribution shown in 
Figure~8a reveals several low-mass cores that formed early, as well as
several high-mass ``peaks'' with high birth ranks.
Another surprising result, for Run A, is the final mass of the most massive
core, which is 35\% of the total mass of the system. This core, the
first one to form, clearly experienced runaway accretion. By contrast, the
most massive core formed in any simulation of KB has a mass of
order $6-8\%$ of the total mass of the system. It is important to
understand these various features. For the rest of \S5.4, we 
will discuss the results from Run A; Runs B and C will be discussed in
\S 5.5.

\subsubsection{Early Ejections}

The presence of
cores in the bottom left corner of Figure~8a is easily explained. Some cores
that form early can be ejected after experiencing close encounters
with other cores. Once a core is ejected from the dense structure where it
originally formed, it finds itself in a low-density region where there is
little gas to accrete. The mass growth of that core then levels off at
a constant value, a process discussed by KB. To illustrate this, we focus
on 3 early-forming cores that end up with very low masses, 
cores~\#8, \#10, and \#14 (crosses
in Fig.~8a). Figure~9a shows the distribution of cores in the 
system at $t=1.084$, just before core~\#10 is ejected. 
The system at that time contains
19 cores, 10 of them forming a dense cluster embedded in a common
gas filament (not shown). This cluster, indicated by the small square,
is enlarged and displayed in Figure~9b. Figure~9c shows the trajectories
of these 10 cores, from the locations where they formed (open circles)
to their locations
at $t=1.154$ (solid circles). Several cores are ejected, including
the cores~\#8, \#10, and \#14.

Figure 9d shows the mass evolution of these 3 cores. After formation,
their masses grow rapidly by accretion, until they are ejected.
After being ejected, cores \#10 and \#14 no longer accrete gas, and remain
at constant mass throughout the reminder of the simulation. Core \#10 is
ejected at very large velocity (see the quasi-straight trajectory in
Fig.~9c), and moves only through low-density regions thereafter.
Core \#14 remains bound to the main cluster, but ends up orbiting the
cluster at a ``safe'' distance, never coming close enough to accrete
gas from the cluster's envelope. Core \#8 is ejected at low velocity, turns
back, and falls into a dense clump at $t=1.541$. Its mass then increases
slightly by accretion. At $t=1.904$, that clump, which still contains
core \#8, merges with the main cluster. The mass of core \#8 increases again
by accreting gas from the main cluster, before being ejected a second time,
after which its mass remains constant.

\begin{figure}[t]
\vskip-.3in
\hskip.8in
\includegraphics[width=4.7in]{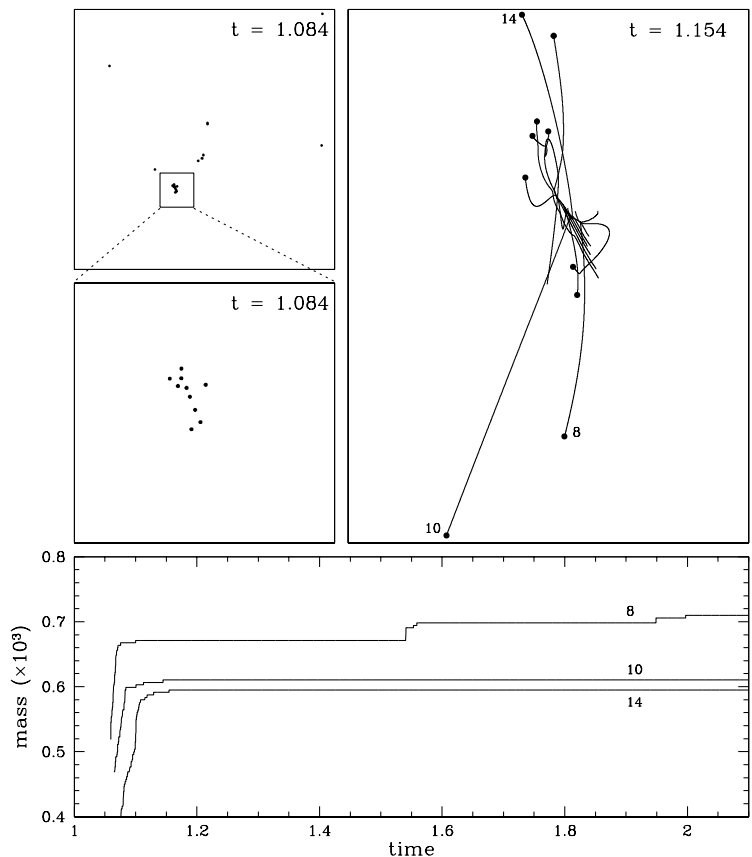}
\caption{(a, top left panel) snapshot of the distribution of cores
in the computational volume at $t=1.084$. The system contains a total
of 19 cores; (b, middle left panel) enlargement of the region indicated by a
square in the previous panels, showing a dense clusters composed of
10 cores; (c, top right panel) the same cluster is shown at
$t=1.154$ (solid dots), along with the trajectories of the cores
between $t=0$ and $t=1.154$ (curves). The cores \#8, \#10, and \#14,
which are ejected, are indicated;
(d, bottom panel) time evolution of the mass of cores \#8, \#10, and \#14.}
\end{figure}

\subsubsection{Local Competitive Accretion}

To understand the origin of the massive cores
at high birth rank, we need to consider
the nature of competitive accretion between cores. 
This process was described in detail by \citet{bbcp97,bbcp01}.
There are 4 basic
arguments for why cores that form earlier should reach higher masses:
(1)~since they form earlier, they have more time to grow by accretion,
(2)~the cores that form early will deplete their surroundings by accretion,
reducing the amount of gas available to cores that form later, (3)~by
being more massive, the cores that form early have a stronger gravitational
potential than the ones forming late, making them more efficient in
accreting the reminder of the surrounding gas, and (4) if several cores of
different masses are present inside a gaseous clump, the most massive
cores will tend to reside in the center of the clump where there
is presumably more gas to accrete. While argument~(1) is general, 
arguments~(2), (3), and (4) are valid only if cores form out of the
same clump, and are therefore competing for the same
surrounding gas. If the final cluster forms by the merging of dense clumps,
and if cores form in these clumps prior to the final merging, then
there will be competitive accretion within each clump, but
not across clumps. The very first core that forms in a particular
clump can then grow by accretion and reach a high mass, no matter
how late core formation in that clump started.

\begin{figure}[t]
\hskip0.8in
\includegraphics[width=4.8in]{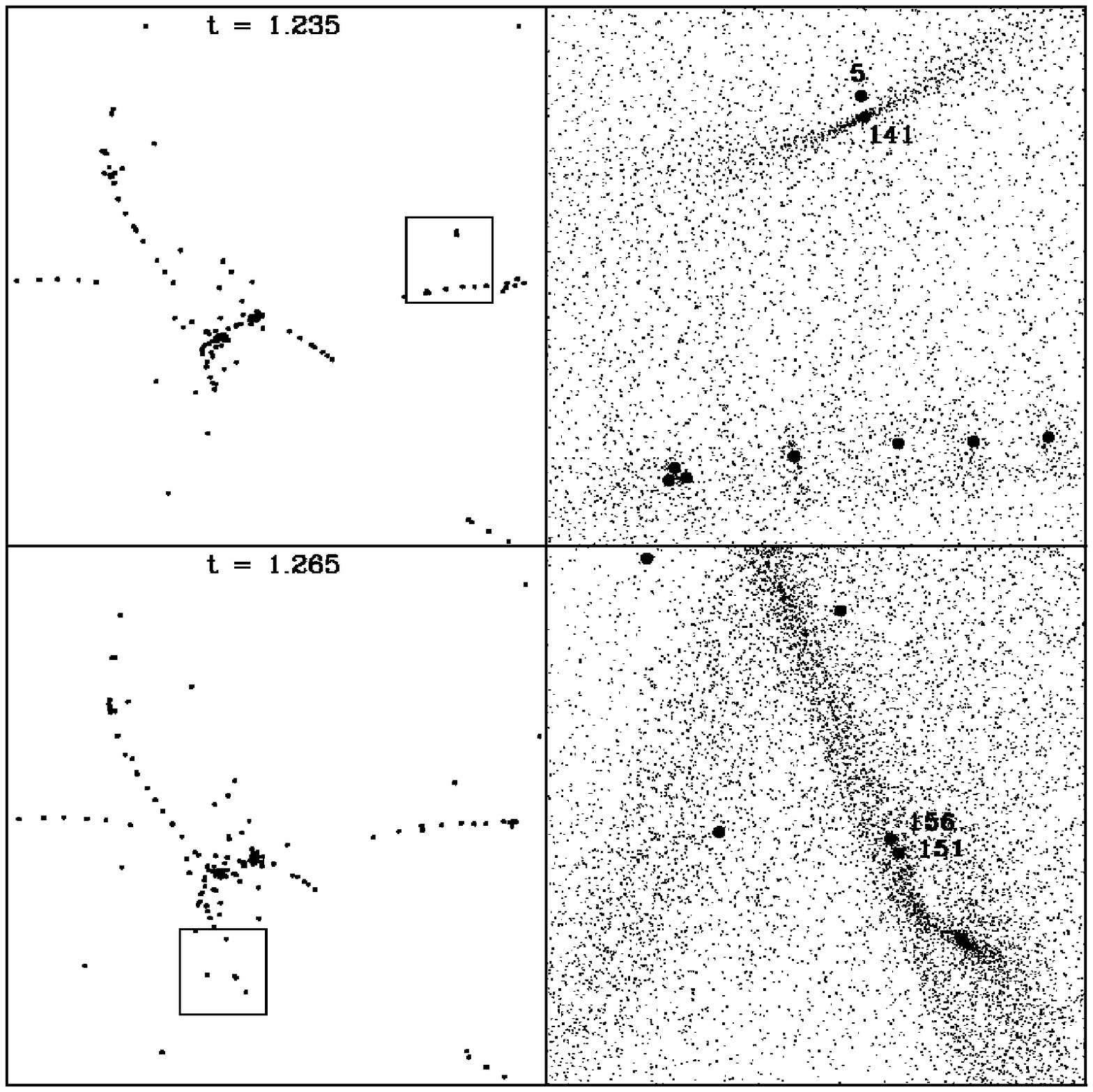}
\vskip0.0in
\caption{(a, top left panel) distribution of cores at $t=1.235$,
immediately after core \#141 formed; (b, top right panel) enlargement of
the region indicated by a square in the previous panel, showing both gas 
particles (small dots) and cores (large dots). Cores \#5 and 141
are indicated by numbers; (c, bottom left panel) and (d, bottom right panel)
similar to (a) and (b), but at time
$t=1.265$, immediately after core \#156 formed.}
\end{figure}

To illustrate that, we focus on 3 cores, cores~\#141, \#151, and \#156, which
are located in two late ``peaks'' (asterisks in Fig.~8a). The top panels
in Figure~10 shows the system at $t=1.235$, immediately after the
formation of core~\#141. As we see, that core did not form inside the
main cluster, but inside an emerging filament located away from the
main cluster, and was the very first core to form there (core~\#5 is
a fast-moving core that was ejected early and is seen in projection).
The bottom panels in Figure~10 show the system at $t=1.265$, immediately after
the formation of core~\#156. Again, cores~\#151 and \#156 formed in a filament
away from the main cluster, and were the first two cores to form there.
Local competitive accretion enables these cores to grow and reach a
high mass before they fall into the main cluster.

\subsubsection{Violent Infall and Late Starburst}

\begin{figure}[t]
\vskip-0.3in
\hskip0.4in
\includegraphics[width=5.6in]{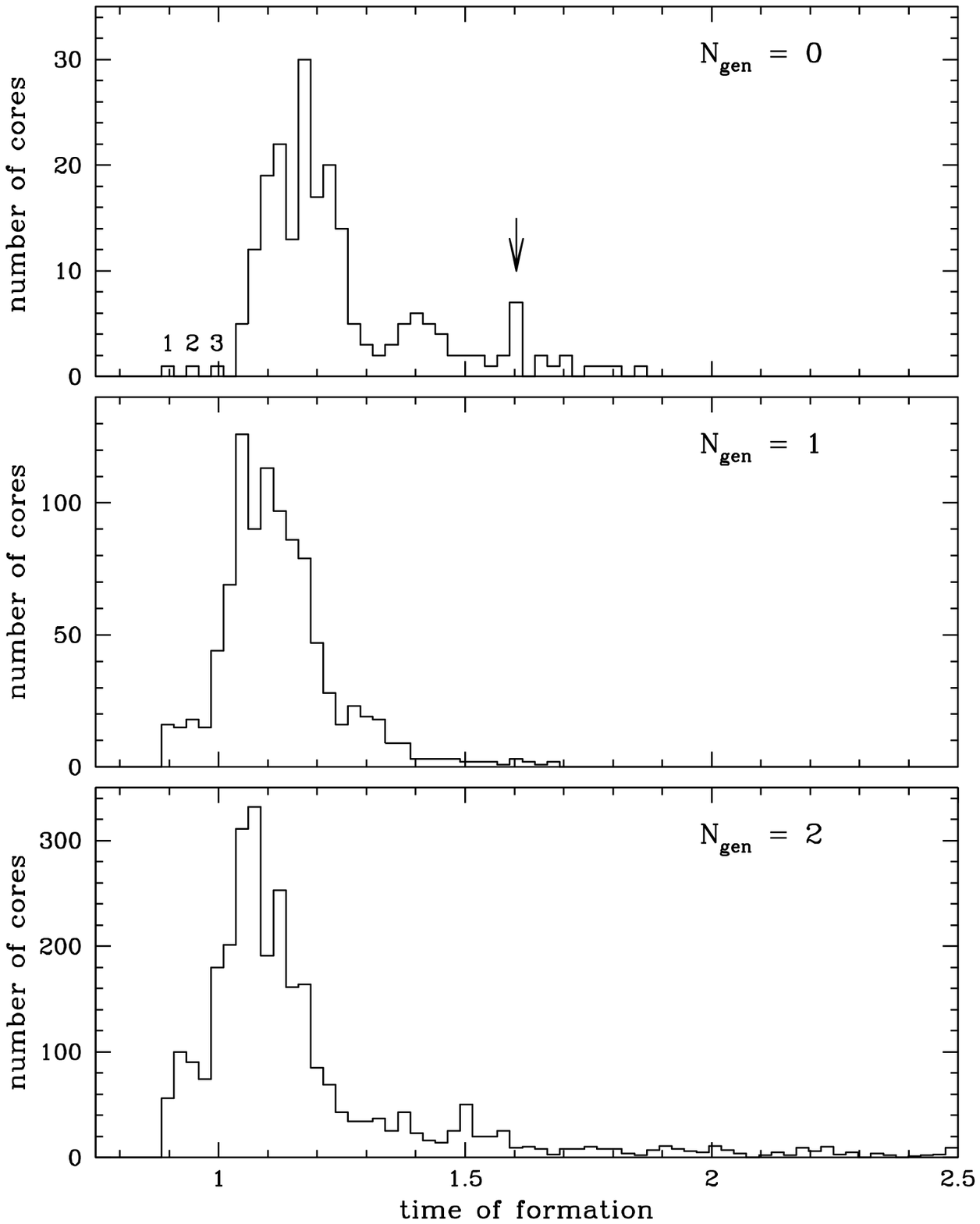}
\vskip-0.2in
\caption{Histogram of the number of protostellar cores
versus their epoch of formation.
(a, top panel) Run A ($N_{\rm gen}=0$);
(b, middle panel) Run B ($N_{\rm gen}=1$);
(c, bottom panel) Run C ($N_{\rm gen}=2$). The arrow in the top panel
identifies the late starburst. The first three cores formed in Run~A
are also identified.
}
\end{figure}

There is another peak in Figure~8a that we wish to consider. This peak
contains cores with birth ranks 199--203 (solid circles in Fig.~8a), 
that formed very late. 
These 5 cores formed almost simultaneously, between $t=1.595$ and
$t=1.609$, all inside a small region of diameter 0.006 located
{\it right in the middle of the cluster}.
Figure~11 shows a histogram of the formation time of cores. 
The formation of cores \#199--203 in Run~A appears as a burst,
which is indicated by an arrow
in the top panel.
We need to understand how this burst occurred and how these cores managed
to reach a high mass in a such a crowded environment. Clearly, we cannot
invoke local competitive accretion here, since these cores, unlike
cores \#141, \#151, and \#156, formed in a region that already contained
many other cores that were already significantly more massive.

Consider the state of the system at time $t\sim1.5-1.6$
(see $8^{\rm th}$ and $9^{\rm th}$ panels in Fig.~3). At that stage,
there is still a substantial amount of gas located far from the main cluster.
Some of that gas, which is located in dense filaments, will form new
cores, that will later fall into the main cluster. The remainder of the
gas, which is located between the filaments, does not reach high enough
densities to trigger the formation of cores. It will therefore
remain in the form of gas until it falls into the main cluster. Because
that gas comes from large distances, it gets accelerated over a long period
of time, and therefore falls inside the main cluster at large velocities.
This large velocity reduces the effectiveness of gravitational focusing,
making it more difficult for that gas to accrete onto
the cores already present in the cluster. 
Instead, that gas is rapidly decelerated by a strong shock and gets compressed
to a very high density, which triggers the formation of several
cores in a burst.

\subsubsection{Runaway Accretion}

\begin{deluxetable}{cccccc}
\tabletypesize{\footnotesize}
\tablecaption{Mass of the First Cores ($\times10^3$).\label{tbl-3}}
\tablewidth{0pt}
\tablehead{
\colhead{Mass} & \colhead{$t_1=0.9010$} & \colhead{$t_2=0.9522$} & 
\colhead{$t_3=1.0055$} & \colhead{$t_4=1.0356$} & \colhead{$t_5=1.0361$}
}
\startdata
$m_1$ & 0.435    & 9.205    &   21.942 & 31.693   & 31.925 \\
$m_2$ & $\cdots$ & 0.405    &    3.922 &  4.494   &  4.494 \\
$m_3$ & $\cdots$ & $\cdots$ &    0.401 &  4.166   &  4.242 \\
$m_4$ & $\cdots$ & $\cdots$ & $\cdots$ &  0.412   &  0.435 \\
$m_5$ & $\cdots$ & $\cdots$ & $\cdots$ & $\cdots$ &  0.381 \\
\enddata
\end{deluxetable}

To understand the very large mass of core~\#1, we need to consider the
formation history of the first few cores. Looking again at Figure~11a,
we immediately notice something special about the first three cores.
These cores formed at well-separated times, after which core formation
proceeded very rapidly.
Table~3 shows, for the first 5 cores, the quantity $m_i(t_j)$, defined
as the mass of core~$i$ at the time of formation of core~$j$ (with $j\geq i$).
All cores are created roughly with the same mass (diagonal in Table~3).
However, core~\#2 formed significantly later than core~\#1, and
during the time interval $t_2-t_1$, 
core~\#1 accreted more than 20 times its original mass
(growing from 0.000435 to 0.009205), so that when core~\#2 formed,
core \#1 was already more massive by a factor of 23. The process then repeats
itself: core~\#3 formed significantly later than core~\#2, and during
the time interval $t_3-t_2$,
core~\#1 and \#2 experienced significant growth. By $t=t_3$,
these 3 cores have mass ratios 55:10:1. This process goes on, and
by $t=t_4$, the first 4 cores have mass ratios 77:11:10:1 (core~\#2 is
ejected between $t_3$ and $t_4$, and stops growing afterward).
Finally, this process terminates: core~\#5 is created almost
immediately after core~\#4, providing little time for the latter to grow,
so that at $t=t_5$, cores~\#4 and \#5 have comparable masses. We are then
in the ``big peak'' in Figure 11a, with many cores forming at
comparable times.

\subsection{Formation and Growth of Protostellar Cores; Higher-Resolution 
Simulations}

We have focused so far
on Run A, and investigated the origin of the various
features found in Figure~8a. Figures~8b and 8c show the results for
Runs~B and C, respectively. The results are qualitatively similar to Run~A.
There is a definite tendency of the mass to decrease with increasing
birth rank. The distributions are very wide, however, and show several
peaks at high birth rank, which we again attribute to competitive
accretion. 

\subsubsection{Early Ejections}

There is a noticeable difference in the higher-resolution runs: 
the absence of very-low mass
cores at small birth ranks. Early ejections seem far less common in
these simulations. This can be explained using the following
argument. Consider an envelope of gas of mass $M$ and radius $R$,
containing several cores of mass 
$m$, forming a small, bound cluster of radius $r$, inside which the
mean core spacing is $\Delta$. The gravitational force $f$ between
cores is proportional to $m^2/\Delta^2$,
and therefore the acceleration $a$ of the
cores varies with $m$ as
\begin{equation}
a={f\over m}\propto{m^2/\Delta^2\over m}\propto{m\over\Delta^2}\,.
\end{equation}

\noindent For a fixed cluster radius $r$, the core spacing $\Delta$ depends
on the number of cores, and for a fixed cluster mass, that number of cores
depends on the resolution, such that $\Delta\propto m^{1/3}$
(if the mass per core is smaller, the cluster contains more cores,
thus those cores are closer to one another). Hence
$a\propto m^{1/3}$.
If a core is ejected from the cluster,
its terminal velocity should be of order $v\approx(2ar)^{1/2}\propto m^{1/6}$.
But the escape velocity from the whole system is of order 
$v_{\rm esc}\approx(GM/R)^{1/2}$, independent of $m$. Therefore,
lowering $m$ makes it more difficult for cores ejected from the
cluster to escape the common envelope of gas.

\subsubsection{Local Competitive Accretion and Late Starburst}

Figures 8b and 8c show several peaks, corresponding to high-mass cores
with high birth ranks. We examined several cases, indicated by asterisks.
In all cases, we found that the high masses resulted from local competitive 
accretion. All these cores formed in regions located away from the main
cluster, often in emerging filaments. In several cases, these filaments 
contained a string of cores (as seen, for instance, in Fig.~5), and the
core that reached a high mass was the one located at the very end
of the string. This is an example of local competitive
accretion, and perhaps the core at the end is able to accrete from a
larger solid angle.

We did not find a late starburst (or coreburst) in the high-resolution
simulations. The gas located in very low regions does fall into the
cluster in the form of gas, but tends to be accreted onto the
cores rather than forming new cores. Accretion is much more efficient
in the higher-resolution simulations simply because of the sheer
number of cores in the cluster at late time. That number is very
small for Run~A because (1) core~\#1 contains more than half the final
mass of the cluster, and (2) as we shall see in \S5.6, the cluster
suffers a great deal of evaporation at late time in Run~A. Hence, the cross
section for accretion onto cluster cores is greatly reduced in the 
low-resolution simulation compared with the high-resolution ones,
which explains why the late starbust phenomenon is seen only at
low resolution.

\subsubsection{Runaway Accretion}

We did not observe the kind of runaway accretion experienced by the
first core in Run A. While that core reaches a final mass of 0.358
(35.8\% of the total mass in the computational volume), the final masses
of the first cores formed in Runs B and C are 0.00295 and 0.000803,
respectively.
Actually, the core which ends up with the largest mass in Run C is
not core \#1, but rather core \#707, with a final mass of 0.00263.
In Run A, runaway accretion occurred because the first core formed 
significantly earlier than the others. With higher resolution, several 
lower-mass cores form almost simultaneously inside the first region that
reaches the threshold density, and competitive accretion
among these cores prevents runaway accretion.

Comparing all these results with the ones presented in \S5.4, we conclude
that local competitive accretion is a fundamental process that occurs at
all resolutions, except in situations where a small number of cores
results in the formation of a single cluster (as in the simulations of KB).
In such a case, competitive accretion does occur, but it is not local. 
Other phenomena, like early ejection, late starburst,
and runaway accretion, appear to be peculiarities of low-resolution
runs, and become less common as the resolution increases.

\subsection{The Protostellar Cluster}

All three simulations end up forming a dense cluster of protostellar
cores. In this section, we study the assembly history and structure
of that cluster.

\subsubsection{Mass and Cluster Members}

\begin{figure}[t]
\vskip-0.4in
\hskip0.5in
\includegraphics[width=5.2in]{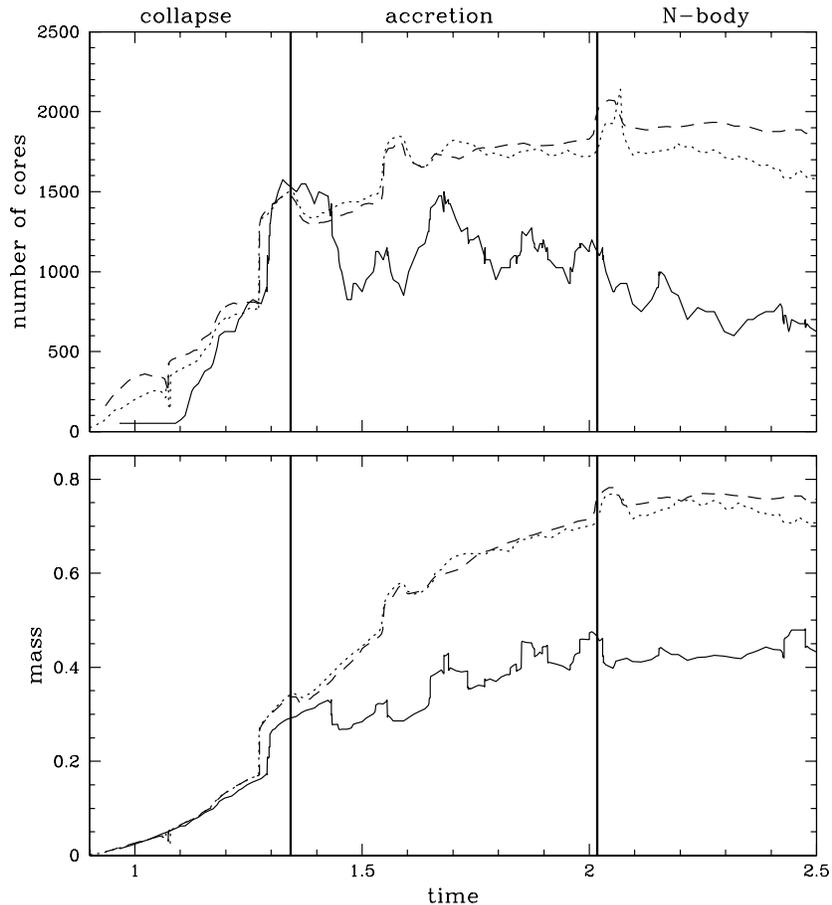}
\vskip-0.2in
\caption{(a, top panel) evolution of the number of cores inside the
main cluster; (b, bottom panel) evolution of the mass in cores inside
the main cluster. On both panels, the three curves correspond to
Run A ($N_{\rm gen}=0$, solid curve),
Run B ($N_{\rm gen}=1$, dotted curve),
and Run C ($N_{\rm gen}=2$, dashed curve). On the top panel, the number
of cores
for Runs A and B have been multiplied by factors of 25 and 2.75, respectively,
to allow a better comparison.
The thick vertical
lines indicate the transitions between the collapse, accretion,
and N-body regimes, as indicated.}
\end{figure}

To identify the cores that belong to the cluster,
we first use visual inspection to make an initial
estimate $({\bf r}_{\rm cl})_{\rm init}$ of the center of the cluster.
Then, using an iterative method, we find a self-consistent solution
for the center of mass ${\bf r}_{\rm cl}$ and radius $r_{200}$ of
the cluster, such that (1) the center of mass of the cores located inside
a sphere of radius $r_{200}$ centered on ${\bf r}_{\rm cl}$ is indeed
${\bf r}_{\rm cl}$, and (2) the mean density $\rho_{200}$
inside that sphere is 200 times the mean density in the
system. Notice that these densities are computed using the cores only,
without the gas, as in KB.

Figure~12a shows the time-evolution of the number of cores in the cluster, 
for all three runs. We have rescaled the number of cores for Runs A and
B by factors of 25 and 2.75, respectively, for comparison with Run C. 
The number of cores for Run A (solid curve) varies tremendously, in a
somewhat chaotic way. These large fluctuations can be attributed to small
statistics, Run A having the smallest total
number of cores. However, this
is not the only explanation. Our method for determining cluster membership
assumes spherical symmetry, but in Run A, the cluster tends to be triaxial
during most of the calculation, and also
the sphere of radius $r_{200}$ which
contains the cluster members tends to follow the motion of core \#1, since
that core contains most of the mass of the cluster. So as core \#1
experiences brownian motion around
the center of the cluster, cores located near the surface of the sphere
keep ``falling'' in and out of the cluster. This does not occur with the
higher-resolution runs, because the cluster tends to be quite spherical,
and no single core dominates its mass.

During the collapse regime ($t<1.34$), cores are forming inside the
cluster, increasing the number of cluster members. There are few ejections
during this phase, and the net effect is a near-monotonic increase in
the number of cores. Once the system enters the accretion regime, few
new cores are forming, and the competing processes are ejections from
the cluster and accretion of cores that formed in secondary clumps
outside the main cluster. Indeed, the main cluster experiences a major
merger with another, comparable cluster, at $t\sim1.29$, near the end
of the collapse phase, resulting in a sudden increase in the number of cores.
For Runs B and C, the number of cores tends to increase during the
accretion regime, indicating that accretion of cores formed in
subclumps dominates over ejections. For Run A, however, the ejections
tend to dominate. Once the system enters the N-body regime
($t>2.02$), the
number of cores steadily decreases for Runs A and B. All the cores formed
outside the main cluster have been accreted along with the remaining gas,
and ejections become the dominant process. However, the number of cores
in Run C remains nearly constant, indicating that very few
ejections are taking place.

Figure 12b shows the evolution of the mass of the cluster. The results are 
nearly identical for all three runs up to $t\sim1.29$, when the major
merger takes place. {From} that point, Run A differs significantly from the
other two runs, as the mass of the cluster grows much more slowly. 
Interestingly, in the N-body regime, the mass of the cluster, 
for Run A, remains nearly
constant even though the number of cores drops significantly. This indicates
that only low-mass cores are ejected at late time.

The most striking result shown in Figure 12b is the similarity between
the results of Runs B and C. 
The curves are essentially indistinguishable
up to the beginning of the accretion phase, and
then remain very similar up to the end of the simulation, having
in particular the same local maxima at $t=1.33$, 1.58, and 2.05,
and the same local minima at $t=1.37$ and 1.61. Notice also the
sudden increase in mass at $t=1.54$, corresponding to a merger
with a smaller subcluster of cores. This merger does not
occur in the low-resolution simulation.

\subsubsection{The Density Profile}

\begin{figure}[t]
\vskip-0.4in
\hskip0.0in
\includegraphics[width=5.3in]{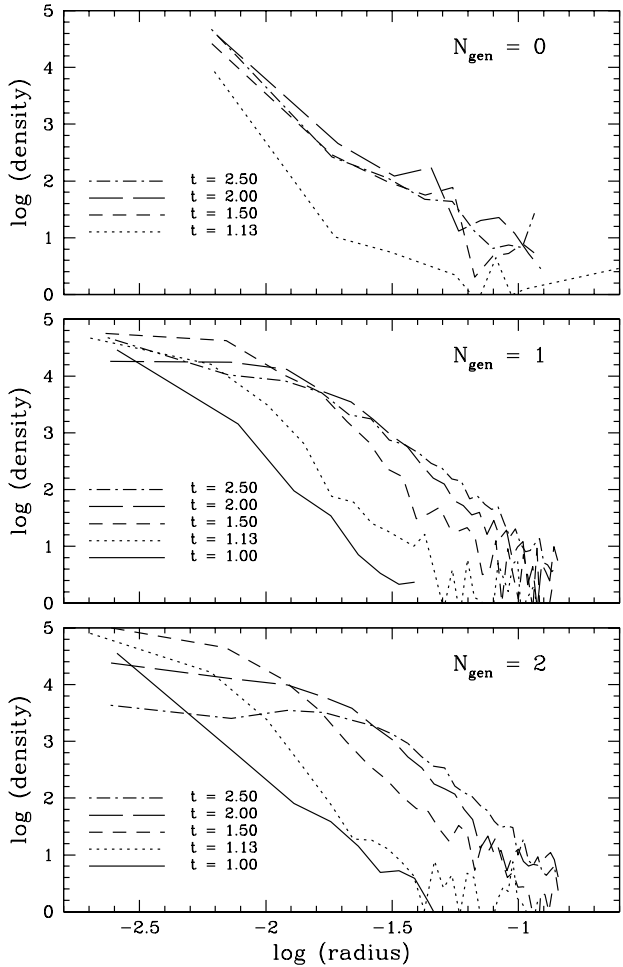}
\caption{Density profile of the main cluster, for Runs A (top panel),
B (middle panel), and C (bottom panel). The various curves correspond
to different times, from bottom to top: 
$t=1.00$ (bottom curves), 1.13, 1.50, 2.00, and 2.50.
There is no curve plotted for $t=1.00$ in the top panel
because the cluster contained only two cores at that time in Run~A.}
\end{figure}

We computed the density profile of the cluster by adding up the mass
of the cores inside radial bins. Figure~13 shows the evolution of
the profile, for all three runs. For Run~A, the profile is very steep at
the center, simply because core \#1, which undergoes runaway accretion,
totally dominates the mass of the cluster. For Runs~B and C, the cluster
profile starts up roughly as a power law, but soon acquires a core/halo
structure, as in the simulations of KB. In these two runs, the central density
drops and the profile flattens at late time ($t>2$), corresponding to the
epoch where the total mass of the cluster stops growing (see Fig.~12).
At these late times. the central part of the cluster inflates, probably as
a result of gravitational heating by tight binary cores.

\begin{figure}[t]
\vskip-0.4in
\hskip0.0in
\includegraphics[width=5.3in]{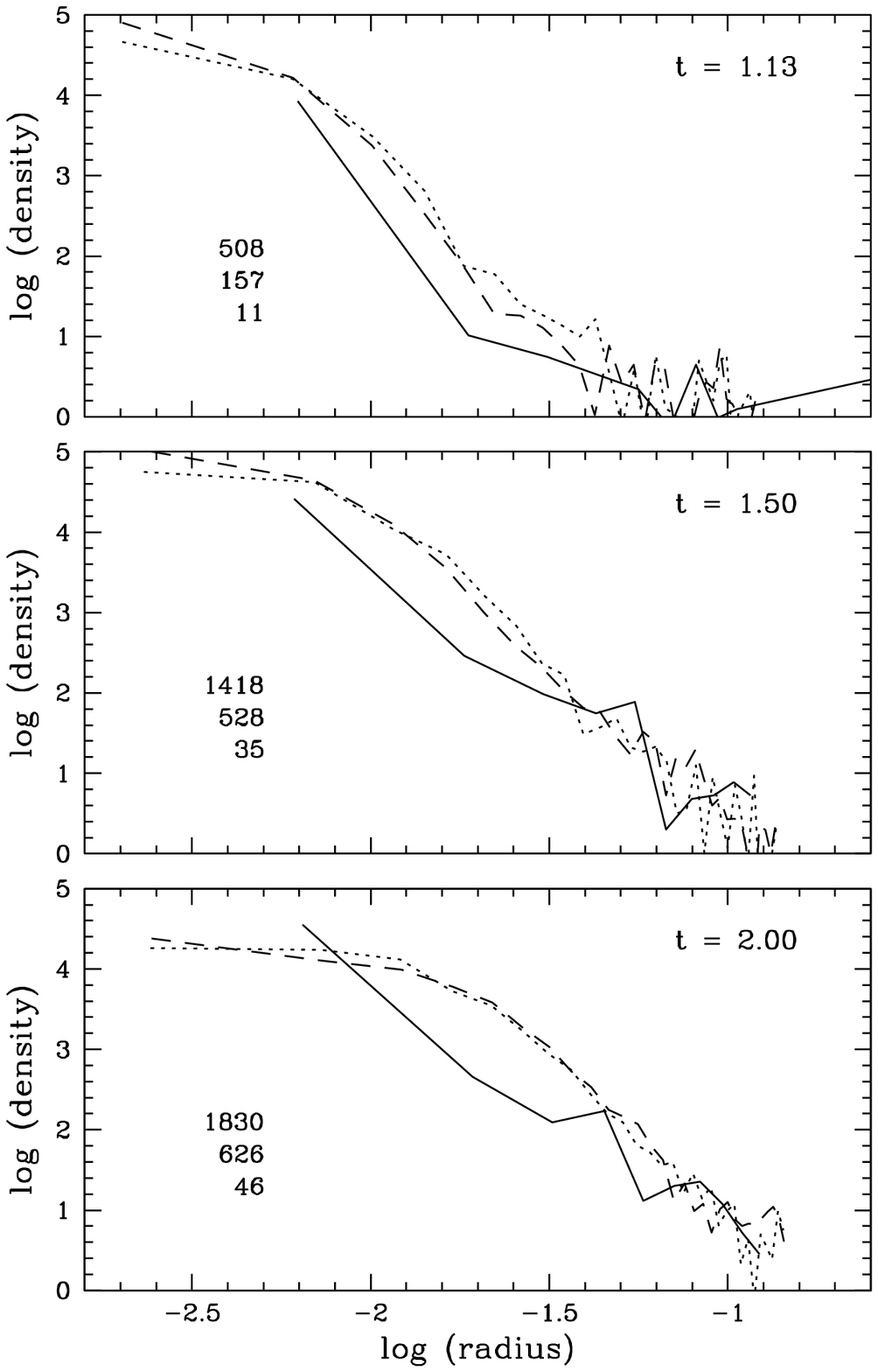}
\caption{Density profile of the main cluster at times
$t=1.13$, 1.50, and 2.00, as indicated.
The solid, dotted, and dashed curves correspond to Runs A, B, and C,
respectively. The numbers in the bottom left corner of each
panel indicate the number of cores in the cluster at these times
(top number: Run~C; middle number: Run~B; bottom number: Run~A).}
\end{figure}

Figure~14 shows a comparison among the three runs of the cluster 
profile at some particular times ($t=1.13$, 1.5, and 2). The density
profiles found in Runs~B and~C are very similar, at all epochs. This
is quite striking, considering that the number of cores in the cluster
roughly quadruples between $t=1.13$ and $t=2$, while the mass in the
cluster increases by a factor of 9. The profile for Run A is totally
different. Most of the mass of the cluster in contained in core \#1,
located in the center of the cluster.

\subsection{Initial Mass Function of Protostellar Cores}

\begin{figure}[t]
\vskip-0.5in
\hskip0.2in
\includegraphics[width=6.0in]{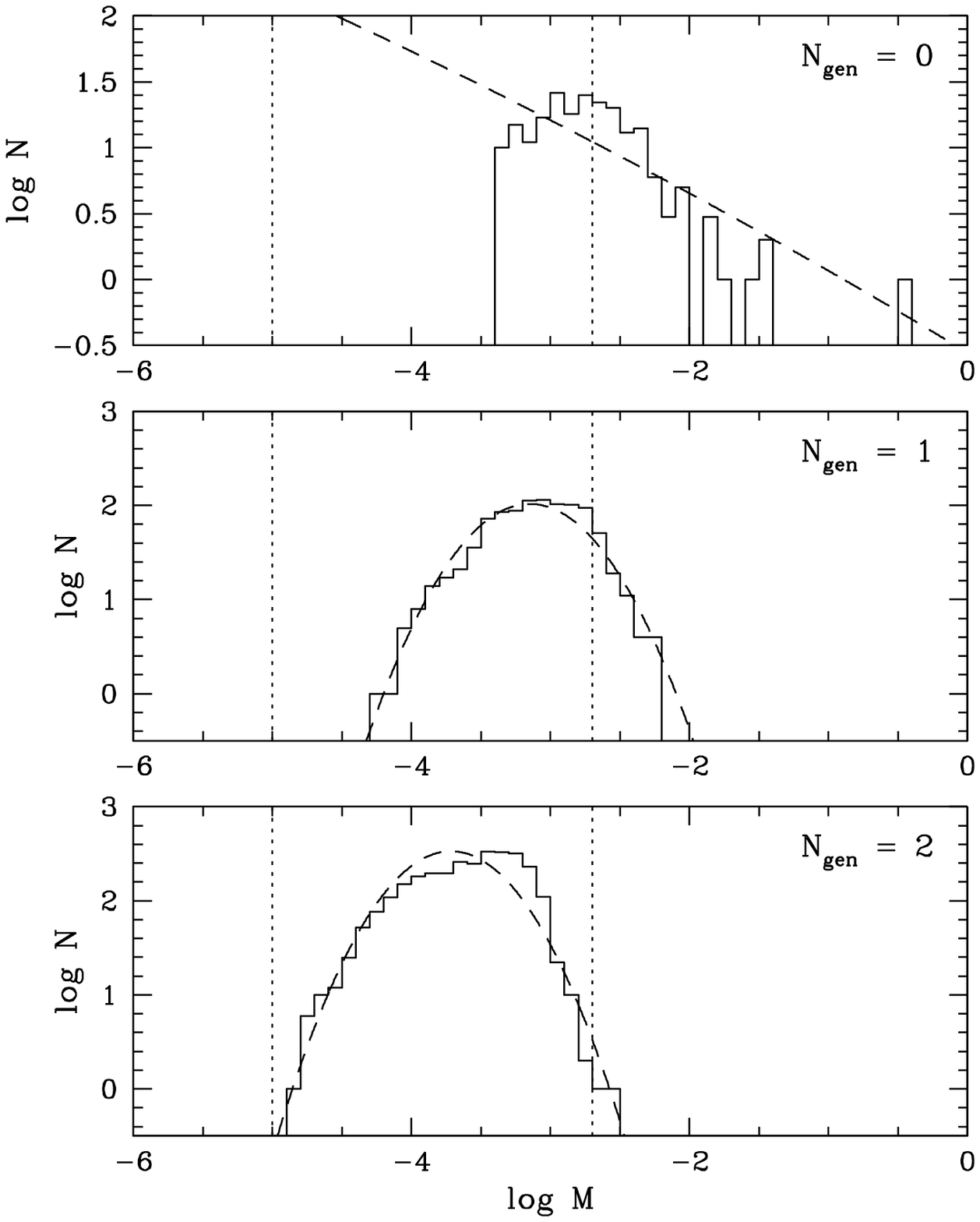}
\vskip-0.5in
\caption{Mass distribution of protostellar cores.
(a, top panel) Run A ($N_{\rm gen}=0$);
(b, middle panel) Run B ($N_{\rm gen}=1$);
(c, bottom panel) Run C ($N_{\rm gen}=2$).
On each panel, the left and right dotted lines indicate the
Jeans mass at densities $\rho=\rho_c$ and $\rho=\bar\rho$, respectively.
The dashed curves show the results of a least-square fit to a log-normal
distribution.}
\end{figure}

Figure 15 shows the final mass distribution of protostellar cores
at $t=2.5$, when nearly all the gas has been accreted.
For Run~A, the existence of a core containing 35\% of the
total mass of the system results in a skewed distribution. 
Interestingly, the simulations of \citet{tp04} show similar features
(Fig.~17 in their paper), suggesting that runaway accretion is
happening in their simulations as well.
For Runs~B and 
C, the distributions are roughly log-normal,
with some skewness toward high masses, as the dashed curves show.
KB reported that the mass distribution peaks at the initial Jeans
mass of the system. Our results are consistent with that claim, but
only for Run~A. As the number of splitting generations, or equivalently
the mass resolution of the algorithm, increases, the distribution shifts
to lower masses while keeping the same width in $\log M$. Clearly,
it is the resolution, and not the initial Jeans mass, that determines
the location of the peak. As the resolution increases, the algorithm
can form cores with smaller masses, until the resolution is so high that 
the Jeans mass $(M_J)_c$ at the threshold density $\rho=\rho_c$
is resolved.
Hence, increasing the resolution moves the low-mass end of the distribution
to lower masses. This leads to the formation of a larger number
of cores, which then have to compete for accretion. As a result,
the high-mass end of the distribution also moves to lower masses.
Note that the shift to lower masses results only from increasing $N_{\rm gen}$
because $\rho_c$ is the same in all three runs.

These log-normal distributions (for Runs B and C) are consistent with
the numerical results of KB. However, they are inconsistent with observations
that show a peak at small masses and a power-law like
behavior at the high-mass end. There are several possible explanations for 
this. Once sinks form, they can no longer fragment. In the real world,
these objects might fragment, increasing the number of low-mass
objects relative to high-mass ones (notice that sink merging, which we also
ignore, would have the opposite effect). We did not take turbulence into
account in these simulations. Turbulence could affect the dynamical evolution
of the system differently at different scales, affecting the final shape of
the mass distribution. Probably even more importantly, feedback effects, if
included, could slow down the growth of massive cores by accretion. Finally,
we could consider changing the slope of the 
power spectrum of initial density fluctuation, though KB did this
\citep{kb01}, and found that the distributions of core masses remained
log-normal.

\section{CONVERGENCE}

The three simulations we have performed all start with $64^3$ particles,
and use identical initial conditions. However, particle splitting increases
the effective resolution of the simulations, which is $128^3$ for Run B
and $256^3$ for Run C. Since the initial conditions are identical but 
the effective resolution varies, this set of simulations constitutes a 
convergence study, which can be used to estimate the minimum resolution
necessary to obtain reliable results.

It is clear, however, that some results simply will not converge. By
requiring that sink particles are created with enough gas particles
to insure that they are resolved (that is, Scenario II, by opposition
to Scenario I), the initial mass of sinks depends on resolution, and
as a result the IMF of cores shown in Figure~15 shifts to lower
masses as the resolution increases. However, the 
results of Run C have converged, since that simulation does resolve the 
Jeans mass. If we added a Run D with $N_{\rm gen}=3$ to our set of simulations,
the results of Runs C and D would be identical, because a third level
of splitting would never occur: before the density gets large enough to
make particles split a third time, these particles would turn into sinks
(see Fig.~2). This convergence is {\it numerical\/} in the sense that
numerical parameters like $\rho_c$ and $N_{\rm gen}$ determine the solution
that the simulation converges to. A solution that would converge
{\it physically\/} does not exist, and therefore cannot be achieved,
in a system with an isothermal equation of state, because no physical
process limits the minimum mass of cores. In the real universe, the
assumption of isothermality breaks down at high densities when the
gas becomes optically thick, and that in turns leads to a physical
minimum mass for cores.

Looking at the macroscopic properties of the final cluster of cores, it is
clear that the results of Runs A and B are significantly different, while
the results of Runs B and C are very similar, indicating that convergence
has been achieved. In particular, the mass history of the cluster (bottom
panel of Fig.~12) and the density profile at various times (Fig.~14) are
strikingly similar for Runs B and C. In Run A, the first core formed
underwent runaway accretion, which affected the further evolution of
the cluster. No such runaway accretion occurred in Runs B and C. We
believe that the likelihood of such occurrence is reduced as the
resolution increases, because as more cores are formed, the time interval
between the formation of a core and the next one is reduced, thus
increasing the competition for accretion.

Finally, there are other results, arguably less interesting, that 
show convergence. In particular, the formation time histograms shown in
Figure~11 are quite similar for Runs B and C, and different for Run A, with
the bulk of the cores forming at later time. 

\section{ILLUSTRATIVE EXAMPLES}

To compare the results to observations, it is convenient to use the
relations in \S4.3 to convert the densities, masses, etc. to physical
units. Because the simulations contain 500 Jeans masses initially, 
the total mass
and other properties are uniquely fixed by a choice of temperature and density.
We set the temperature at $10\,{\rm K}$
for all these examples because this temperature
is characteristic of both dust and gas temperatures in well shielded regions
before stars form \citep{leung75,evansetal01,youngetal04}. 
We will consider below constraints imposed by the assumption of isothermality.
Observers commonly use total particle density 
[$n=n({\rm H_2})+n({\rm He})$], where $\rho=\mu_n m_H n$,
with $m_H$ the atomic mass unit. 
For a fully molecular cloud with 25\% helium by mass, $\mu_n = 2.29$.
We will use $n$ for the initial density.

\begin{deluxetable}{cccccccc} 
\label{phystab}
\tabletypesize{\footnotesize}
\tablecaption{Physical Quantities}
\tablewidth{0pt}
\tablehead{
\colhead{Case} & \colhead{$n$} & \colhead{$\rho$} & \colhead{$M_{\rm tot}$} & 
\colhead{$L_{\rm box}$} & \colhead{$t_{\rm dyn}$} & \colhead{$\tau$} &
\colhead{$\tau_{\rm kern}$} \cr
\colhead{} & \colhead{(cm$^{-3}$)} & \colhead{(g cm$^{-3}$}) & 
\colhead{(\msun)} &
\colhead{(pc)} & \colhead{(yr)} & \colhead{(neper)} & \colhead{(neper)}
}
\startdata
1&$10^2$ & 3.8\ee{-22} & 1.0\ee4 & 12.1 & 6.28\ee6 & 6.9\ee{-4} & 1.9\ee{-2}\cr
2&$10^3$ & 3.8\ee{-21} & 3.2\ee3 & 3.8 & 1.99\ee6 & 2.2\ee{-3} & 6.0\ee{-2}\cr
3&$10^4$ & 3.8\ee{-20} & 1.0\ee3 & 1.21 & 6.28\ee5 & 6.9\ee{-3} & 1.9\ee{-1}\cr
4&$10^5$ & 3.8\ee{-19} & 3.2\ee2 & 0.38 & 1.99\ee5 & 2.2\ee{-2} & 6.0\ee{-1}\cr
\enddata
\end{deluxetable}

With equation~(\ref{mjinit}) for the Jeans mass, we have in physical units:
$M_{J,{\rm init}}=6.33T^{1.5}n^{-0.5}\msun=200 n^{-0.5}\msun$ for
$T=10\,{\rm K}$.
It follows that $M_{{\rm tot}} = 1.0\ee5 n^{-0.5}\msun$,
and the size of the region, 
$L_{{\rm box}}=(M_{{\rm tot}}/\bar\rho)^{1/3}=121n^{-0.5}{\rm pc}$. 
The dynamical time, 
$t_{\rm dyn}=(G \bar\rho)^{-1/2}=6.3\ee7 n^{-0.5}{\rm yr}$. 
Values for these quantities are given in Table~4
for different values of $n$.

The assumption that the gas remains isothermal depends on its ability to
cool. In dense regions, the gas cools by collisions with dust grains,
which radiate in a continuum \citep{goldreich74,doty97,youngetal04}.
To remain isothermal, the optical depth in the continuum near the peak
emission wavelength should be less than unity, measured from the center
of the region to the edge.  The initial optical depth is computed from
$\tau = \kappa \bar\rho L_{\rm box}/2$, where $\kappa$ is the opacity of
dust per gram of gas.
Emission from dust at 10 K peaks at a wavelength around 350 \micron.
Calculations of dust opacities for dust that has coagulated and acquired
ice mantles, as may be expected in dense regions, have been done by
\citet{ossenkopf94}.
Observations are generally well matched
by the opacities from column 5 in their table, known as OH5 opacities.
The value for 350 \micron, assuming a gas to dust ratio of 100, is
$0.1\,{\rm cm}^2{\rm g}^{-1}$ of gas. The values of $\tau$ in Table~4
are computed from these assumptions. 

Because we form sinks at a density $f_{\rm sink} = 4\ee4$ 
times the initial density (using the notation of \S3.1),
we must check that the region around the sink is optically thin just
before sink formation. A convenient measure for this is the optical
depth calculated for the radius of the SPH kernel at $\rho_c = f_{\rm sink}
\bar\rho$. The radius of the kernel is about 3 times the local particle
spacing $\Delta r$. The particle spacing is constrained by
$\Delta r\leq0.5L_{\rm box}f_{\rm sink}^{-1/3}n_{\rm part}^{-1}
2^{-N_{\rm gen}}$, where $n_{\rm part}$ is the initial number of
particles along the edge of the computational volume
(64 for these simulations). 
Thus, we have for the center to edge optical depth
of a kernel,
$\tau_{\rm kern}=\tau F$,
where
$F=6f_{\rm sink}^{2/3}n_{\rm part}^{-1}2^{-N_{\rm gen}}$.
For $N_{\rm gen} = 0$, 1, 2, $F = 109.6$, 54.8, 27.4.
Particle splitting helps to keep the optical depth in a kernel from
rising too far above the initial value, since increasing $N_{\rm gen}$
lowers $F$.
Note that $\tau_{\rm kern} \propto \tau \propto n^{0.5}$ because
of the scalings compelled by assuming that we have 500 Jeans masses.
With the criterion that $\tau_{\rm kern} < 1$, these calculations would
be valid up to $n \sim 10^5$ cm$^{-3}$, so we do not show any entries
in Table~4 with $n$ above that value. With $f_{\rm sink}=4\ee4$, 
the maximum density would be about $4 \times 10^{9}{\rm cm}^{-3}$.
Alternatively, we could use the accretion radius ($r_{\rm acc}$, see \S 2.4)
instead of the kernel radius. These two radii turn out to be very similar, so 
the results are not significantly affected.

The 4 cases listed in Table~4 are simply scaled to different
values of the initial density. The first is roughly similar to the
example given by \citet{kb00},
which they compare to the
Taurus cloud. Our case 1 has about twice the mass of that of KB because
we have roughly twice as many Jeans masses. Our definition of Jeans mass
is slightly different, and other differences combine to make our value
for $L_{\rm box}$ about twice their value. The cases with higher initial
densities approximate the conditions in cluster forming regions.
Case 3 is similar to the conditions in the L1688 (Ophiuchus) cluster,
which has $500 - 1000 \msun$ in a region extending about 1 pc 
\citep{johnstoneetal00}.
Case 4 approximates conditions in the  massive, dense regions studied 
by \citet{shirleyetal03a};
the median radius of gas traced by CS emission
in their study is 0.37 pc, and the mean virial mass is about 1200 \msun.
Mass estimates based on the 350 \micron\ emission \citep{muelleretal02}
for a subsample of those sources yield mean values around 300 \msun.

Putting the mass functions shown in Figure~15
into physical units is probably premature
because we have not yet included many effects, such as feedback from
early star formation. If we proceed with these caveats in mind, we find
that the mass function, in the simulation with $N_{\rm gen} = 2$, and
for Case 3, peaks around 0.3 \msun, with a maximum mass around
2.0 \msun\ and a minimum mass of 0.03 \msun, well into the region of 
brown dwarfs. Case 4 produces even lower mass objects, with a peak
in the distribution around 0.1 \msun. This is not surprising since
there is no limit to how small the Jeans mass may get with an isothermal
equation of state. Clearly the other effects,
which are the subject of future work, will need to be included before
these distributions can be compared to the observed mass functions.

\section{DISCUSSION}

We have performed three simulations of fragmenting molecular clouds,
using identical initial conditions, but different levels of particle
splitting. Each additional level corresponds to an increase by
a factor of 2 in length resolution and 8 in mass resolution. We have
discovered several interesting phenomena, which we described in \S5
above. Some of the results, such as the runaway accretion
and the late starburst, were found only in the lowest-resolution
simulation, and thus are clearly resolution-dependent. Such phenomena
are peculiarities that happen sometimes when the number of cores is
relatively small.
We found three results that are not resolution-dependent: (1)
the existence of four distinct regimes in the evolution of the cloud,
(2) the phenomenon of local competitive accretion, and
(3) the tight relationship between the mass range of the
IMF and the resolution limit of the algorithm, until the calculation
is fully resolved.

In these initial simulations, we have neglected several physical processes,
such as turbulence, non-isothermality, and feedback, which will be addressed
in future work. We believe that the three main results stated above are
robust, and will remain valid once we include additional physical processes
(though this remains to be proven). First, the existence of the four
distinct regimes should not be affected by additional physics. The
gas will always end up into cores given enough time, so there will always
be a N-body regime at the late stages. The fact that half the
gas turns into cores during the growth regime, while the other half
is accreted during the accretion regime, is most likely a consequence of
the Gaussianity of the initial conditions.

Local competitive accretion occurs because the timescale for the
fragmentation of subclumps into cores is shorter than the timescale
for these clumps to merge and form the final cluster. This could possibly
be affected by turbulence or feedback, since these processes could
delay core formation inside subclumps. In particular, the feedback
from the first core forming in a given subclump could possibly prevent
the formation of other cores in the same subclump, so that the final cluster
would form by the merger of several subclumps, each one containing only
one core. It remains to be seen if the effect of feedback could be
that drastic. 

We need a particular mass or length scale in the problem
to determine the mass range of the IMF. In our simulations, the
gas is isothermal, and there are no physical scale in the system. It is
then a numerical scale, the minimum Jeans mass resolved by the algorithm, that
determines the lower edge of the IMF. In simulations with a barotropic
equation of state, the scale corresponds to the Jeans mass at the
density where the gas becomes adiabatic. Turbulence and feedback could
introduce a physical scale as well. But in all cases, there is
no apparent reason for having a relationship between the mean initial Jeans
mass and the lower edge of the IMF. Whatever the
initial Jeans mass is, we expect the cloud to fragment down to the
lowest mass scale allowed by either the physical processes or the numerical
resolution.

\section{SUMMARY AND CONCLUSION}

This paper presented simulations of the fragmentation of a
molecular cloud in which artificial fragmentation is prevented by using
particle splitting, a technique that had not yet been applied to this
kind of problem. With this technique, we can follow the evolution of the 
entire cloud, 
from the initial conditions to the formation of a star cluster, while
fully resolving the Jeans mass at all densities.

The main objective of this paper was to demonstrate the feasibility of
simulating the formation of a star cluster by cloud fragmentation, while
properly resolving the Jeans mass throughout the entire simulation. We have
successfully shown that, with the use of particle splitting, this can be 
achieved at a small fraction of the computational cost of a
standard high-resolution or a ``zoom-in'' simulation. In particular,
our largest simulation (Run C), which fully resolves the Jeans mass,
has an effective resolution of $256^3$, or 17 million particles,
while the actual number of particles in that simulation varies but never 
exceeds 1.1 million. The gain in performance is huge, and would only
increase with additional splitting generations.

We have identified four distinct phases in the evolution of the cloud,
corresponding to four different regimes. Initially, the cloud is in the
growth regime. Overdense regions become denser and eventually
form a network of filaments. Then, in the collapse regime, the gas in
dense regions is converted into protostellar cores. In the accretion regime,
the remaining gas, which started up in underdense regions, is mainly accreted
by the existing cores. Finally, in the N-body regime, most of the gas has 
disappeared, and the evolution of the cluster is governed by N-body dynamics.
These various regimes were certainly present in previous simulations, such
as the ones of KB, but were not explicitly identified. The existence of
a collapse and an accretion phase, and the fact that roughly 50\% of the
gas is removed during the collapse phase and 50\% during the accretion
phase, is most likely a mere consequence of the Gaussianity of the initial
conditions, though this remains to be tested.

In the lowest-resolution simulation, we have noticed several interesting 
phenomena, such as early ejection, local competitive accretion, late starburst,
and runaway accretion. Early ejections (also noticed by KB and
\citealt{bbb02b}) occur during
few-body encounters, and explain how cores that form early stop accreting and
end up having a low mass. Local competitive accretion is a new phenomenon
that we have identified. It occurs when several clumps of gas fragment to
form sub-clusters of cores that later merge to form the final cluster. The
first core formed in each clump does not compete for accretion until
other cores form in the same clump, and can therefore
reach high masses even if
it formed late in the overall simulation. This explain most of the
high-mass peaks at high birth ranks seen in Figure~8. Late starburst is caused
by gas that was never dense enough to form cores until it falls into the main
cluster at late times and gets suddenly shocked to very high densities.
Runaway accretion occurs when the first core formed in the simulation
formed significantly earlier that the next cores, and thus gets an
early start in accreting gas. The mass difference then increases as a result of
competitive accretion, and that core ends up containing a large fraction
(up to one third) of the mass of the entire system. While all these phenomena
were observed in the lowest-resolution simulation, only local competitive 
accretion was observed in the higher-resolution simulations. This suggests
that local competitive accretion is a fundamental process that greatly
affects the evolution of the system, while the other phenomena are less
fundamental, and might not have occurred at all (or to a very different
extent) if we had used different initial conditions. We believe
that it is the small number of cores formed in the lowest-resolution
simulation, and not its inability to resolve the Jeans mass, that is 
responsible for the occurrence of these phenomena. 

The final distribution of the core masses are roughly log-normal (in
agreement with the isothermal
simulations of KB), except for the lowest-resolution
simulation, where the runaway accretion of core \#1 results in a very
skewed distribution. The location of the distribution shifts to
lower masses as the resolution increases, until the resolution is
sufficient to resolve the Jeans mass. This is a consequence of our
decision to follow ``Scenario~II'' for the simulations with
$N_{\rm gen}=1$ and 2, requiring that dense clumps are converted
into cores only if they contain at least $n_{\min}$ particles. Had
we followed Scenario~I, all simulations would have produced
(presumably) distributions with the same mean. We found that the
mean of the distribution is determined entirely by the resolution, and
not by the mean Jeans mass at the initial time, in contradiction
with the claim of KB.

\acknowledgments

This work benefited from stimulating discussions with A. Burkert, R. Fischer,
R. Klessen, C. Matzner, J. Scalo, A. Urban, and A. Whitworth. We thank
the referee for very helpful comments.
All calculations were performed at the Texas Advanced Computing Center.
We are pleased to acknowledge the support of 
NASA Grants NAG5-10825, NAG5-10826, and NAG5-13271. H.M. thanks the
Canada Research Chair program and NSERC for support.

\appendix

\section{INITIAL CONDITIONS}

The technique used for generating initial conditions is a generalization
of the Zel'dovich approximation commonly used for cosmological simulations.
The same technique was used by KB, but their description lacks details.
In this appendix, we provide a more detailed description. Notice that most
of this derivation can be found in \S10.2 of Coles \& Lucchin (1995).

To set up initial conditions, we lay down $N$ equal-mass particles on a uniform
cubic lattice inside the computational volume, and displace these particles to
represent the density perturbation 
as a Gaussian random field with a particular power spectrum. We
then adjust the particle velocities by requiring that the perturbation
is growing with time. First, we start with equations~(1), (2), and~(4). 
We eliminate $\rho$ using 
$\rho=\bar\rho(1+\delta)$, where $\delta$ is the density contrast.
Then, assuming that the perturbation is initially small, we 
neglect the pressure gradient, and only
keep terms that are linear in $\delta$, ${\bf v}$, and $\phi$. 
Equations~(1), (2), and (4) reduce to
\begin{eqnarray}
&&{\partial\delta\over\partial t}+\nabla\cdot{\bf v}=0\,,\\
&&{\partial{\bf v}\over\partial t}=-\nabla\phi\,,\\
&&\nabla^2\phi=4\pi G\bar\rho\delta\,.
\end{eqnarray}

\noindent To solve these equations, we first take the divergence of 
equation~(A2). This introduces a term in $\nabla\cdot{\bf v}$, which we
eliminate using equation~(A1), and a term in $\nabla^2\phi$, which we
eliminate using equation~(A3). We get
\begin{equation}
{\partial^2\delta\over\partial t^2}=4\pi G\bar\rho\delta=\omega^2\delta\,,
\end{equation}

\noindent where
$\omega\equiv(4\pi G\bar\rho)^{1/2}$. The general solution is
\begin{equation}
\delta({\bf r},t)=A({\bf r})e^{\omega t}+B({\bf r})e^{-\omega t}\,.
\end{equation}

\noindent We assume that at the initial time $t_i$, the perturbation is
in a pure growing mode, and therefore $B=0$. 
Equation~(A5) reduces to
\begin{equation}
\delta({\bf r},t_i)=A({\bf r})e^{\omega t_i}\,.
\end{equation}

\noindent 
To solve for the velocity, we decompose the density contrast and
velocity into sums of plane waves,
\begin{eqnarray}
\delta({\bf r},t_i)&=&
e^{\omega t_i}\sum_{\bf k}A_{\bf k}e^{-2\pi i{\bf k}\cdot{\bf r}}\,,\\
{\bf v}({\bf r},t_i)&=&
\sum_{\bf k}{\bf v}_{\bf k}e^{-2\pi i{\bf k}\cdot{\bf r}}\,,
\end{eqnarray}

\noindent 
where the sums are over all plane waves that satisfy the periodic boundary
conditions and whose frequency does not exceed the Nyquist
frequency,
\begin{equation}
{\bf k}=(n_1,n_2,n_3)/L_{\rm box}, \qquad 
n_1,n_2,n_3=0,1,2,\ldots n_{\rm nyq}\,,
\end{equation}

\noindent where $n_{\rm nyq}=N^{1/3}/2$.
We substitute equations~(A7) and (A8) into equation~(A1),
which becomes
\begin{equation}
\omega e^{\omega t_i}\sum_{\bf k}A_{\bf k}e^{-2\pi i{\bf k}\cdot{\bf r}}
=2\pi i\sum_{\bf k}{\bf k}\cdot{\bf v}_{\bf k}
e^{-2\pi i{\bf k}\cdot{\bf r}}
\,.
\end{equation}

\noindent Since the plane waves are orthogonal functions, equation~(A10)
implies
\begin{equation}
\omega e^{\omega t_i}A_{\bf k}=2\pi i{\bf k}\cdot{\bf v}_{\bf k}\,.
\end{equation}

\noindent To solve for ${\bf v}_{\bf k}$, we make the assumption
that the velocity field is vorticity-free. This implies that
${\bf k}\times{\bf v}_{\bf k}=0$, and therefore 
${\bf v}_{\bf k}=|{\bf v}_{\bf k}|{\bf k}/k$ and
${\bf k}\cdot{\bf v}_{\bf k}=|{\bf v}_{\bf k}|k$, where $k=|{\bf k}|$.
Equation~(A11) then gives
\begin{equation}
{\bf v}_{\bf k}=-{i\omega A_{\bf k}{\bf k}\over2\pi k^2}e^{\omega t_i}\,,
\end{equation}

\noindent and equation~(A8) becomes
\begin{equation}
{\bf v}({\bf r},t_i)=-{i\omega e^{\omega t_i}\over2\pi}
\sum_{\bf k}{A_{\bf k}{\bf k}\over k^2}e^{-2\pi i{\bf k}\cdot{\bf r}}\,.
\end{equation}

\noindent 
To compute the particle displacements $(\Delta{\bf r})_i$, 
we integrate the velocity
between times $-\infty$ and $t_i$,
\begin{equation}
(\Delta{\bf r})_i=-{ie^{\omega t_i}\over2\pi}
\sum_{\bf k}{A_{\bf k}{\bf k}\over k^2}e^{-2\pi i{\bf k}\cdot{\bf r}}
={{\bf v}({\bf r},t_i)\over\omega}\,.
\end{equation}

\noindent For a given set of complex amplitudes $A_{\bf k}$, the
displacements and velocities can be computed using equation~(A13) and~(A14).
The amplitudes can be written as
\begin{equation}
A_{\bf k}=|A_{\bf k}|e^{i\phi_{\bf k}}\,,
\end{equation}

\noindent where $\phi_{\bf k}$ is a random phase with uniform probability
between 0 and $2\pi$, and $|A_{\bf k}|$ is related to the power spectrum by
\begin{equation}
P(k)=|A_{\bf k}|^2\,.
\end{equation}

\noindent Notice that the phases must satisfy the 
condition $\phi_{-{\bf k}}=-\phi_{\bf k}$ for $\Delta{\bf r}$ and $\bf v$ 
to be real. We assume a power spectrum of the form
\begin{equation}
P(k)=ck^{-n}\,,
\end{equation}

\noindent where $c$ is a constant. The displacements and velocity depends
upon $c$ and $t_i$ only through the quantity $c^{1/2}e^{\omega t_i}$. This
quantity is arbitrary, but must be chosen sufficiently small that
$\delta({\bf r},t_i)\ll1$, otherwise this linear treatment is not 
valid. As KB point out, it must not be chosen too small either, otherwise the
early evolution of the system would proceed very slowly, resulting in
an excessive amount of CPU time just to get to the nonlinear regime.
We generated initial conditions by imposing that the largest particle
initial displacements $(\Delta{\bf r})_i$ are equal to 10\% of the
mean particle spacing. This limits the initial values of
$\delta({\bf r},t_i)$ to be in the range
$-0.3\leq\delta({\bf r},t_i)\leq0.3$.

%

\end{document}